\documentclass[twocolumn]{aastex631}
\shorttitle{Tracing High-z Outflows with NUV Absorption Lines}
\shortauthors{Kehoe et al.}
\graphicspath{{./}}

\begin{document}

\title{The AURORA Survey: Tracing Galactic Outflows at $z\gtrsim2.5$ with \textit{JWST}/NIRSpec NUV Absorption Lines}

\author{Emily Kehoe}\affiliation{Department of Physics \& Astronomy, University of California, Los Angeles, 430 Portola Plaza, Los Angeles, CA 90095, USA}
\email{ekehoe@astro.ucla.edu}

\author[0000-0003-3509-4855]{Alice E. Shapley}\affiliation{Department of Physics \& Astronomy, University of California, Los Angeles, 430 Portola Plaza, Los Angeles, CA 90095, USA}

\author[0000-0003-4792-9119]{Ryan L. Sanders}\affiliation{Department of Physics and Astronomy, University of Kentucky, 505 Rose Street, Lexington, KY 40506, USA}

\author[0000-0001-9687-4973]{Naveen A. Reddy}\affiliation{Department of Physics \& Astronomy, University of California, Riverside, 900 University Avenue, Riverside, CA 92521, USA}

\author{Natalie Lam}\affiliation{Department of Physics \& Astronomy, University of California, Los Angeles, 430 Portola Plaza, Los Angeles, CA 90095, USA}

\author[0000-0003-1249-6392]{Leonardo Clarke}\affiliation{Department of Physics \& Astronomy, University of California, Los Angeles, 430 Portola Plaza, Los Angeles, CA 90095, USA}

\author[0000-0002-3736-476X]{Fergus Cullen}\affiliation{Institute for Astronomy, University of Edinburgh, Royal Observatory, Edinburgh, EH9 3HJ, UK}

\author[0000-0001-7782-7071]{Richard S. Ellis}\affiliation{Department of Physics \& Astronomy, University College London. Gower St., London WC1E 6BT, UK}

\author[0000-0003-4264-3381]{N. M. F\"orster Schreiber}\affiliation{Max-Planck-Institut f\"ur extraterrestrische Physik (MPE), Giessenbachstr.1, D-85748 Garching, Germany}

\author[0000-0001-5860-3419]{Tucker Jones}\affiliation{Department of Physics and Astronomy, University of California Davis, 1 Shields Avenue, Davis, CA 95616, USA}

\author[0000-0002-0101-336X]{Ali Ahmad Khostovan}
\affiliation{Department of Physics and Astronomy, University of Kentucky, 505 Rose Street, Lexington, KY 40506, USA}
\affiliation{Laboratory for Multiwavelength Astrophysics, School of Physics and Astronomy, Rochester Institute of Technology, 84 Lomb Memorial Drive, Rochester, NY 14623, USA}

\author[0000-0003-4368-3326]{Derek J. McLeod}\affiliation{Institute for Astronomy, University of Edinburgh, Royal Observatory, Edinburgh, EH9 3HJ, UK}

\author{Ross J. McLure}\affiliation{Institute for Astronomy, University of Edinburgh, Royal Observatory, Edinburgh, EH9 3HJ, UK}

\author[0000-0002-7064-4309]{Desika Narayanan}\affiliation{Department of Astronomy, University of Florida, 211 Bryant Space Sciences Center, Gainesville, FL, USA}

\author[0000-0001-5851-6649]{Pascal Oesch}\affiliation{Department of Astronomy, University of Geneva, Chemin Pegasi 51, 1290 Versoix, Switzerland}

\author[0000-0003-4464-4505]{Anthony J. Pahl}\affiliation{The Observatories of the Carnegie Institution for Science, 813 Santa Barbara Street, Pasadena, CA 91101, USA}



\begin{abstract}
We probe galactic-scale outflows in star-forming galaxies at $z\gtrsim2.5$ drawn from the \textit{JWST}/NIRSpec AURORA program. For the first time, we directly compare outflow properties from the early universe to the present day using near-UV absorption lines. We measure ISM kinematics from Fe\,{\sc ii} and Mg\,{\sc ii} absorption features in 41 and 43 galaxies, respectively, and examine how these kinematics correlate with galaxy properties. We find that galaxies with outflows tend to have higher stellar masses, and that maximum outflow velocities increase with stellar mass, SFR, UV slope $\beta$, $E(B-V)$, and $A_V$. We also find that Mg\,{\sc ii} emission is more common in galaxies with lower masses, higher sSFRs, and less dust. These trends are consistent with those in star-forming galaxies at $z<2$ when using the same outflow tracers, suggesting that the feedback from star formation has played a persistent role in shaping galaxy evolution over cosmic time. We also directly compare near-UV and far-UV features in the same NIRSpec spectrum for a $z=5.19$ galaxy, finding consistent ISM kinematics and demonstrating that different tracers yield comparable measurements. We also detect Na\,D absorption in 10 galaxies, which have higher stellar mass, SFR, and dust attenuation compared to galaxies without Na\,D absorption, which is consistent with $z\sim0$ studies. The broad continuum coverage and sensitivity of NIRSpec will enable future studies with larger samples, allowing for robust tests of these trends across a wider dynamic range of galaxy properties.

\end{abstract}

\keywords{galaxies: evolution -- galaxies: high-redshift -- galaxies: kinematics and dynamics}


\section{Introduction}

Galactic-scale outflows are fundamental in understanding galaxy evolution, as they regulate the cycle of baryons between galaxies and their environments. These outflows are typically driven by feedback from massive stars, supernovae, and active galactic nuclei (AGN) \citep{1998Silk, 1999Leitherer, 2005Veilleux, 2006Croton, 2011Murray, 2011King}. By transporting energy, momentum, and enriched material from the interstellar medium (ISM), outflows regulate the gas supply for star formation \citep{2005NDiMatteo, 2005Scannapieco, 2006Croton, 2008Hopkins, 2008Somerville, 2011Gabor, 2015Erb, 2017Beckman}, while simultaneously dispersing heavy elements that enrich the circumgalactic medium (CGM) and intergalactic medium (IGM) \citep{1990Heckman, 1999Martin, 2005Martin, 2000Pettini, 2001Pettini, 2003Shapley, 2004Tremonti, 2007Tremonti, 2005Veilleux, 2005Rupke, 2007Dalcanton, 2008Finlator, 2009Weiner, 2010Steidel, 2011Coil, 2014Peeples, 2017Tumlinson}. Therefore, outflows influence the evolution of the star formation rates (SFR) and chemical enrichment processes within and around galaxies \citep{2012Dave, Hopkins2012, 2013Hirschmann, 2013Vogelsberger, 2017Chisholm}. 

Investigating galactic outflows beyond the local universe is vital for understanding the evolution of galaxies across different epochs. Observational studies at redshifts $0.5 \lesssim z \lesssim 1.5$ using rest-frame near-UV (NUV) low-ionization interstellar absorption lines demonstrated that galactic outflows are widespread in star-forming galaxies, with their incidence and strength correlating with properties such as stellar mass ($M_*$) and SFR \citep{2009Weiner, 2010Rubin, Kornei2012, Martin2012, 2014Bordoloi}. These outflows, traced by cool ($T\sim10^4\text{~K}$) blueshifted gas, regulate star formation, expel enriched material, and shape the evolution of galaxies \citep{2009Sato, 2011Coil, 2007Tremonti, 2011Prochaska,2013Bradshaw, 2015Zhu}. At $z\gtrsim 2$, where the cosmic star formation peaks, studies used rest-frame UV spectroscopy and integral field unit observations of rest-frame optical emission lines from ionized gas (e.g., H$\alpha$, [O\,{\sc iii}]) to characterize outflow kinematics and driving mechanisms \citep{2003Shapley,2010Steidel, 2019Schreiber}. These works highlight the role of both stellar feedback and AGN activity in launching outflows, and suggest that outflow properties are shaped by various stellar properties of the galaxy, such as SFR, $M_*$, and star formation rate surface density ($\Sigma_{\text{SFR}}$) \citep{2010Steidel,2012Talia,2012Newman,2019Schreiber,2019Davies,2022Weldon, 2022Calabro, Kehoe2024}. These studies indicate that outflows play a central role in regulating the baryon cycle and shaping galaxy evolution. The precise drivers and redshift evolution of galactic outflows are still not fully understood, requiring further investigation using a range of observational tracers.

Rest-frame far-UV (FUV) (below 2200\,\AA), NUV (2200--3000\,\AA), and optical absorption lines (e.g., \cite{2003Shapley, 2009Weiner, 2010Steidel,  2012Erb, 2022Weldon}), emission features (e.g., \cite{1990Heckman, 2009Shapiro, 2013Liu, 2014Arribas,  2022Concas}), and X-ray observations (e.g., \cite{2009Strickland, 2010Tombesi}) all serve as valuable tracers of galactic outflows. However, at higher redshifts, the set of available diagnostics becomes incomplete due to observational limitations imposed by the Earth's atmosphere, including either a lack of necessary instrumentation in the observed NUV or significant contamination from the sky background in the observed near-IR. Specifically, at $z\sim0.5-1.5$, rest-frame NUV transitions such as Fe\,{\sc ii} and Mg\,{\sc ii} are available in the observed optical \citep{2009Weiner, 2010Rubin, 2011Prochaska, 2012Erb, Kornei2012, Martin2012, Martin2013, Kornei2013, 2014Bordoloi, 2014Rubin, 2018Feltre,  2021Prusinski}, while at $z>2$, rest-frame FUV tracers are used \citep{2003Shapley, 2010Steidel, 2012Talia, 2022Calabro, 2022Weldon, Kehoe2024}. At $z\sim1$, the rest-frame FUV lines fall into the NUV, which is inaccessible from the ground due to the atmospheric cutoff. At $z>2$, the rest-frame NUV lines shift into the red and the near-IR, where a bright and highly wavelength-dependent sky background poses significant challenges for obtaining robust continuum and absorption measurements. Accordingly, the subset of available outflow tracers is strongly redshift dependent.

These constraints have led to variations in how outflow properties are traced across different cosmic epochs, with each tracer sensitive to different gas phases and physical conditions. Both FUV and NUV lines trace cool outflowing gas, but FUV lines probe both low- and high-ionization gas, whereas NUV lines typically trace only low-ionization gas. Furthermore, optical tracers, such as H$\alpha$, trace gas closer to the launching site of outflows, resulting in differences in outflow detection fractions among the same sample of galaxies \citep{Kehoe2024}. At $z\sim1$, observations using rest-frame NUV tracers like Fe\,{\sc ii} and Mg\,{\sc ii} generally found that outflow velocities and equivalent widths (EW) increased with SFR, $M_*$, and $\Sigma_{\text{SFR}}$ \citep{2009Weiner, 2010Rubin, 2012Erb, Martin2012, Kornei2012, 2014Rubin,2021Prusinski}. In contrast, measurements of rest-frame FUV lines at $z\gtrsim2$ reported weaker or no significant correlations between outflow velocity and galaxy properties such as $M_*$ or SFR \citep{2010Steidel, 2022Weldon, 2022Calabro, Kehoe2024}. These discrepancies may reflect differences in sample selection or biases introduced by using different tracers at different redshifts, or real physical differences between galaxies at different stages of the Universe's history. To ensure consistent comparisons across redshifts, it is important to use the same tracers across different cosmic epochs.

With NIRSpec on \textit{JWST}, high-quality near-infrared spectroscopy is now possible without the interference of the bright sky background imposed by Earth's atmosphere, enabling the detection of rest-frame NUV lines at $z\gtrsim2$ for the first time. This advancement allows for the use of consistent rest-frame NUV absorption-line diagnostics across $z\sim 0.5-3$. Utilizing the same tracers across a broad redshift range will allow for robust evolutionary studies. Here we use the Assembly of Ultradeep Rest-optical Observations Revealing Astrophysics
(AURORA) Cycle 1 \textit{JWST}/NIRSpec program \citep{2025Shapley} to measure the outflow velocities of high-redshift galaxies ($z\gtrsim2.5$) by analyzing rest-frame NUV absorption lines redshifted to observed wavelengths of $\lambda \geq 1 \mu\text{m}$. By comparing the galaxies' outflow signatures, we investigate the relationship between outflow velocity and both stellar properties ($M_*$, SFR, specific star formation rate (sSFR), and $\Sigma_{\text{SFR}}$), and dust properties (UV slope $\beta$, $E(B-V)$, and A$_V$). These correlations provide insight into the physical mechanisms of driving outflows, such as whether they are powered primarily by momentum and thermal energy from stellar winds and supernovae or by radiation pressure from intense star formation. Understanding these relationships will allow us to better understand the role outflows play in the evolution of galaxies in the early universe.

In Section \ref{sec:Observations}, we describe our observations, data reduction, and sample for UV absorption-line analysis. In Section \ref{sec:Measurements}, we detail our methodology for measuring outflow velocities, identifying Mg\,{\sc ii} emitters, measuring galaxy properties such as $M_*$, SFR, sSFR, $\Sigma_{\text{SFR}}$, $\beta$, $E(B-V)$, and A$_V$, and constructing composite spectra. In Section \ref{sec:Results}, we search for correlations between outflow velocity and various galaxy properties, and examine how Mg\,{\sc ii} emission varies with these parameters. Lastly, in Section \ref{sec:Discussion} we compare our results to previous studies of both NUV and FUV absorption lines at low and high redshift, examine differences between Mg\,{\sc ii} emitters and non-emitters, and discuss possible Na\,D detections in our sample. Throughout this paper, we adopt a $\Lambda$CDM cosmology with $H_0 = 70 \text{ km s}^{-1}$, $\Omega_m = 0.3$, $\Omega_\Lambda = 0.7$ and the \cite{2003Chabrier} stellar initial mass function (IMF).

\section{Observations, Reductions, and Sample}\label{sec:Observations}

The spectroscopic sample analyzed here is drawn
from the AURORA survey (GO-1914; PI: Shapley).
The observational strategy and target selection for AURORA
are described in \citet{Shapley2024} and \citet{2024Sanders}.
In brief, AURORA consists of two \textit{JWST}/NIRSpec Multi-Shutter Assembly (MSA)
pointings, one in each of the COSMOS and GOODS-N fields in CANDELS \citep{Grogin2011, 2011Koekemoer}.
A total of 97 $z>1.4$ galaxies was targeted on the two AURORA pointings.
The primary targets on the mask were star-forming galaxies at $z=1.4-4.4$
with predicted detections for faint auroral emission lines enabling direct metallicities.
MSA configurations were filled with $z>6$ targets, quiescent galaxies at $z>2$, strong line emitters, and galaxies with photometric redshifts $z>1.5$.
The bulk of the AURORA sample comprises galaxies that are consistent with the star-forming main sequence \citep{2014Speagle}.
Each AURORA pointing was observed with
the G140M/F100LP, G235M/F170LP, and G395M/F290LP grating/filter
combinations. The typical spectral resolutions are $R\sim1250$ for G140M, $R\sim1400$ for G235M, and $R\sim 1500$ for G395M, and the corresponding pixel samplings are 2\,\AA, 4\,\AA, and 10\,\AA, respectively. The exposure times for each grating were
12.3, 8.0, and 4.2 hours, respectively, obtained using a 3-point dither pattern. This dither pattern was repeated 5 times in each of G140M/F100LP and G235M/F170LP, and twice in G395M/F290LP. Observations across the three grating/filters yielded continuous wavelength coverage from $1-5\rm \mu m$ for each target.

As described in detail in \citet{Shapley2024}, \citet{2024Sanders}, and \citet{Topping2025}, NIRSpec MSA data were reduced in two dimensions (2D) and then optimally extracted to one dimension (1D). These spectra were also corrected for slit losses, with careful attention to the spatially-resolved morphologies of target galaxies (Reddy et al. (in prep)). Flux calibration was performed both in a relative sense from grating to grating, and on an absolute scale, the latter with reference to existing photometric measurements.

\begin{figure*}
    \centering
    \includegraphics[width=\linewidth]{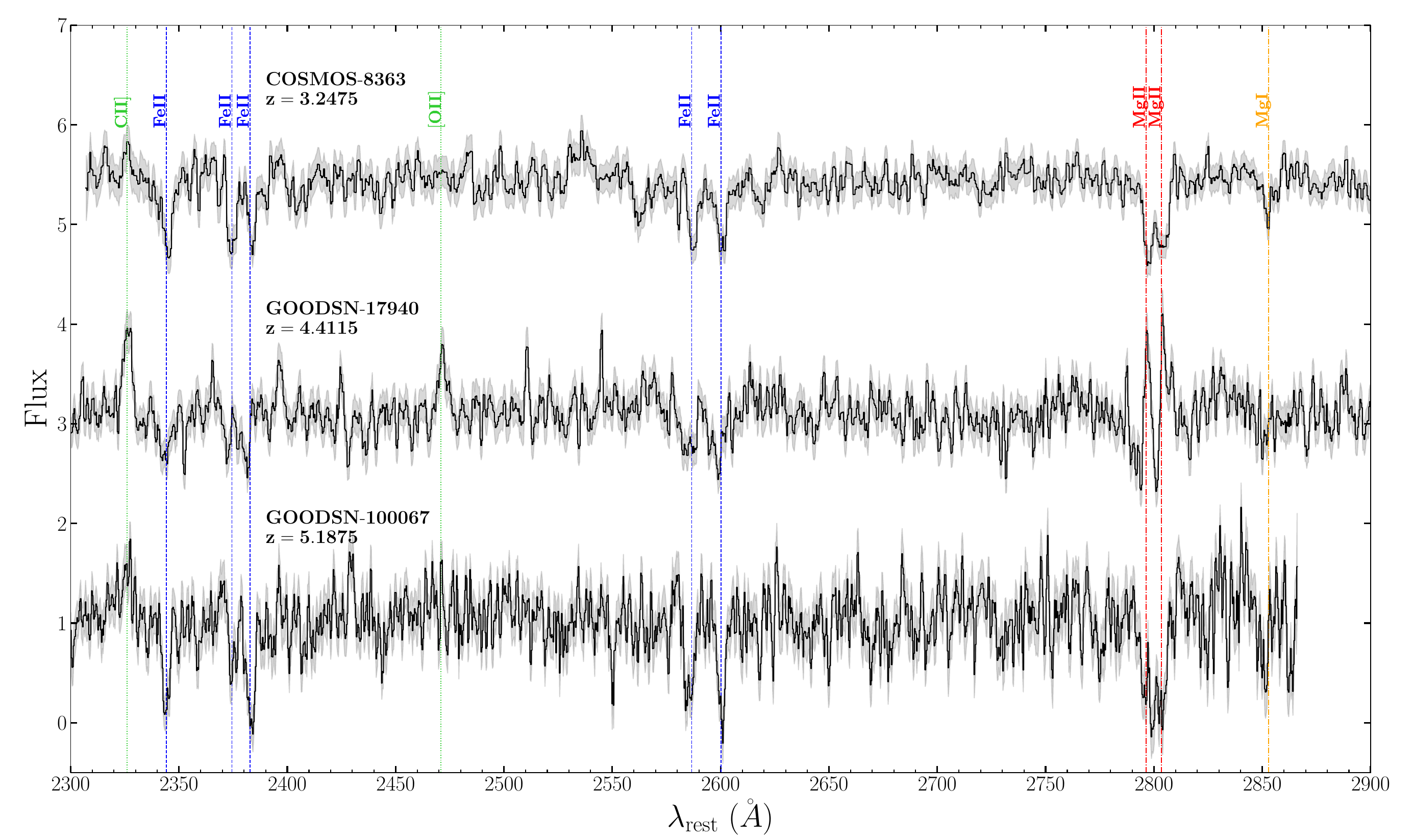}
    \caption{Example of AURORA spectra at 3 different redshifts that are continuum normalized: COSMOS-8363 at $z = 3.247511$ (top), GOODSN-17940 at $z = 4.11541$ (middle), and GOODSN-100067 at $z=5.187539$ (bottom). The error spectra are plotted as gray shaded regions. Blue labels indicate absorption lines that we use to analyze outflow velocities from Fe\,{\sc ii}. Red labels indicate the Mg\,{\sc ii} doublet. Green labels indicate emission lines. The orange label indicates the Mg\,{\sc i} absorption line.}
    \label{fig:1d_spec_compare}
\end{figure*}

\begin{figure*}
    \centering    \includegraphics[width=\linewidth]{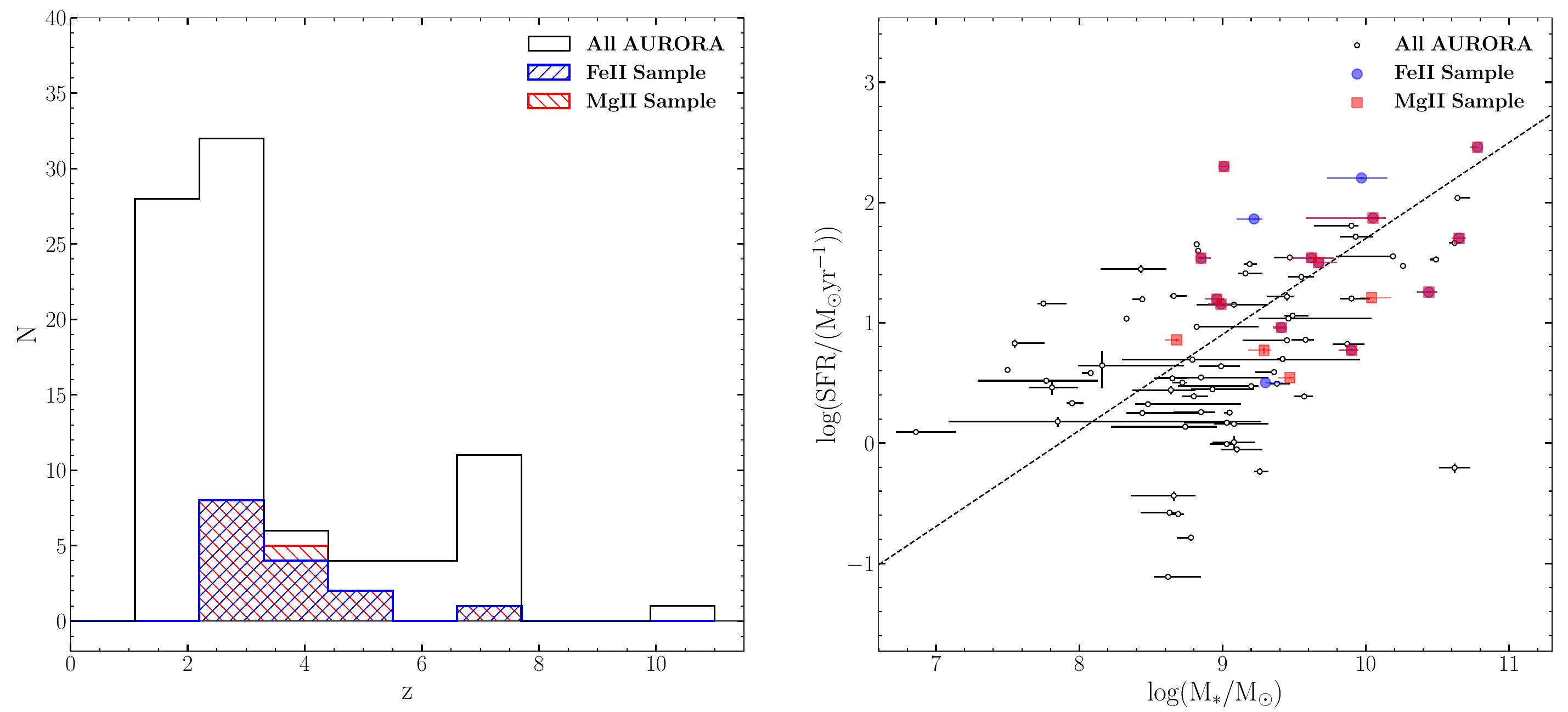}
    \caption{Properties of galaxies from the AURORA survey. Left: Redshift distribution of the full survey (black histogram), of galaxies with measured Fe\,{\sc ii} centroids (blue histogram), and of galaxies with measured Mg\,{\sc ii} centroids (red histogram). Right: The SFR and stellar masses for the full survey (open black circles), galaxies with measured Fe\,{\sc ii} centroids (blue circles), and of galaxies with measured Mg\,{\sc ii} centroids (red squares). The dashed line represents the SFR main sequence from \citet{2014Speagle} at $z\sim 3$. Our Fe\,{\sc ii} and Mg\,{\sc ii} samples scatter well around the star-forming main sequence.}
    \label{fig:sample_demographics}
\end{figure*}

In this work, we analyze galaxies from the AURORA survey with coverage of Fe\,{\sc ii} absorption lines presented in Table \ref{tab:line_windows} and/or Mg\,{\sc ii}$\lambda\lambda 2796,2803$ features presented in Table \ref{tab:line_windows} (see also Figure \ref{fig:1d_spec_compare}). NIRSpec wavelength coverage of these features corresponds to a lower limit in redshift. For Fe\,{\sc ii}, this redshift lower limit corresponds to $z=2.87$, whereas for Mg\,{\sc ii}, the redshift lower limit is $z=2.57$. Galaxies identified as AGN or classified as quiescent are excluded from the sample. Therefore, the total sample of AURORA galaxies with Fe\,{\sc ii} and Mg\,{\sc ii} coverage is 41 and 43, respectively.
The Fe\,{\sc ii} and Mg\,{\sc ii} measurements presented in this work are observed
primarily in the G140M grating. The redshift histogram of the Fe\,{\sc ii} and Mg\,{\sc ii} samples are shown in Figure~\ref{fig:sample_demographics} (left), along with the histogram of the full AURORA spectroscopic sample. Basic galaxy properties (Section 3.3) for the sample are shown in Figure~\ref{fig:sample_demographics} (right), along with the star-forming main sequence out to $z \sim 3$ \citep{2014Speagle}.

\section{Measurements}\label{sec:Measurements}
    \begin{deluxetable}{lccc}
    \tablecaption{Wavelength ranges for line fitting      \label{tab:line_windows}}
    \tablehead{
      \colhead{Line} & \colhead{$\lambda\textsubscript{rest}$ (\AA)} & \colhead{Blue Window (\AA)\tablenotemark{a}} & \colhead{Red Window (\AA)\tablenotemark{a}}
      }
      \startdata
      Al\,{\sc ii} & 1670.79 & $1635-1655$ & $1685-1700$\\
      Fe\,{\sc ii} & 2249.88 & $2200-2200$ & $2300-2330$ \\
      Fe\,{\sc ii} & 2260.78 & $2200-2200$ & $2300-2330$ \\
      Fe\,{\sc ii} & 2344.21 & $2300-2330$ & $2405-2425$ \\
      Fe\,{\sc ii} & 2374.46 & $2300-2330$ & $2405-2425$ \\
      Fe\,{\sc ii} & 2382.77 & $2300-2330$ & $2405-2425$ \\
      Fe\,{\sc ii} & 2586.65 & $2540-2560$ & $2640-2660$ \\
      Fe\,{\sc ii} & 2600.17 & $2540-2560$ & $2640-2660$ \\
      Mg\,{\sc ii} & 2796.35 & $2705-2730$ & $2820-2840$ \\
      Mg\,{\sc ii} & 2803.53 & $2705-2730$ & $2820-2840$ \\
      \enddata
      \tablenotetext{a}{The blue and red regions represent the wavelength ranges used for local continuum fitting for each feature.}
\end{deluxetable}
        \subsection{Fe\,{\sc ii} Kinematics} \label{sec:feii_fits}
            Doppler shifts in low-ionization interstellar absorption lines trace galactic outflows, requiring accurate redshift measurements to calculate the gas velocity relative to the host galaxy. We determined these velocity shifts by measuring the centroids of Fe\,{\sc ii} absorption lines presented in Table \ref{tab:line_windows}. First, we identified significantly detected lines by estimating the line flux using the nonparametric method described in \cite{Kehoe2024}. We then created a composite spectrum from the entire sample to ensure a high S/N. Using this composite spectrum, we identified blue and red continuum regions free of emission and absorption lines to determine the continuum level around each line (Table \ref{tab:line_windows}). We then used the continuum level to measure the EW for each line. We calculated the uncertainties for the flux measurements using Monte Carlo simulations, iterating 1000 times, and identifying lines as significantly detected if their absolute flux values exceed 4.5$\sigma$. This threshold was empirically determined by testing a range of significance levels and selecting the lowest value above 3$\sigma$ that provided robust results upon actual inspection.
            
            For the detected lines, we fit Gaussian curves to measure the observed centroids of the absorption lines. We also calculated the uncertainties for these centroid measurements using Monte Carlo simulations. If the flux from the Gaussian fit exceeded 4$\sigma$, with this threshold determined using the same empirical method as described above, the fit was considered sufficiently robust to utilize the centroid measurement to calculate the outflow velocity. We determined the velocity relative to the systemic redshift using the measured centroids and the spectroscopic redshift from rest-optical emission line measurements of NIRSpec spectra \citep{Shapley2024}, applying the equation:
            
            \begin{equation} \Delta v = \frac{\lambda_{\text{centroid}}-\lambda_{\text{rest}}}{\lambda_{\text{rest}}}\times c \end{equation}
            
           We calculated an inverse-variance-weighted average of the velocity shifts measured from all detected Fe\,{\sc ii} absorption line for each galaxy, denoted as $\Delta v_{\mathrm{Fe_{II}}}$. We measured the velocity for 15 out of 41 galaxies with Fe\,{\sc ii} coverage (37\%), and found an average velocity of $\langle\Delta v_{\mathrm{Fe_{II}}}\rangle = -72\pm26$ km s$^{-1}$. The significantly negative mean velocity is indicative of outflows. Hereafter, we define detected outflows as negative velocity offsets that are greater in magnitude than their uncertainties.
        
        \subsection{Mg\,{\sc ii} Kinematics} \label{sec:mgii_kinematics}
        \subsubsection{Mg \,{\sc ii} Fitting Methodology}
        \label{sec:mgii_fits}
\begin{figure}
                \centering
                \includegraphics[width=\linewidth]{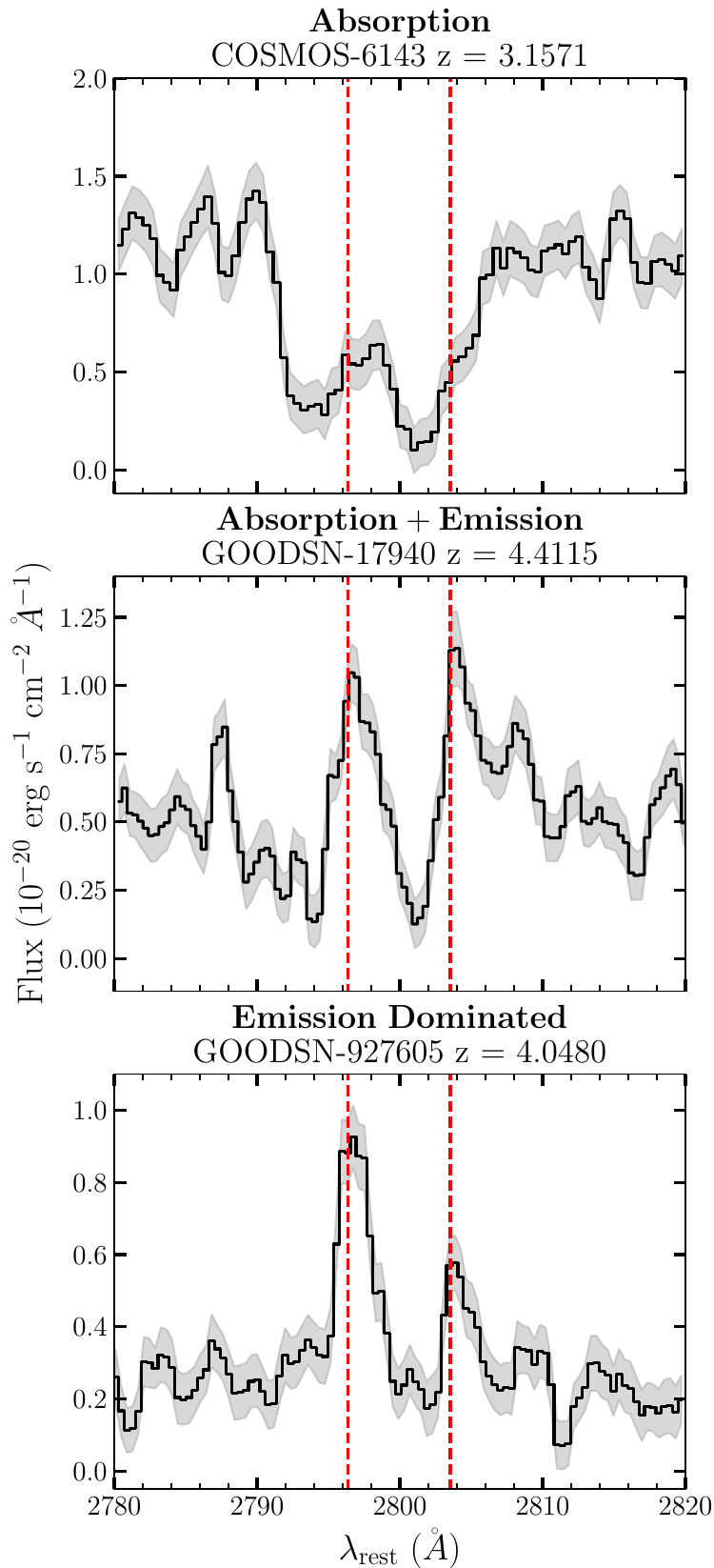}
                \caption{Close up Mg\,{\sc ii} profiles of absorption only (top), both absorption and emission (middle), and emission dominated (bottom). The dashed red lines indicate the rest frame wavelengths of the Mg\,{\sc ii}$\lambda2796$ and Mg\,{\sc ii}$\lambda2803$ lines. The gray shaded region represents the error spectrum.}
                \label{fig:mgii_profiles}
            \end{figure}

            Measuring the centroid from Mg\,{\sc ii} presents a more complex case than measuring the centroids from the Fe\,{\sc ii} lines as the Mg\,{\sc ii} doublet profile profile can appear in emission, absorption, or a combination of both (Figure \ref{fig:mgii_profiles}). As a result, applying a simple Gaussian fit to Mg\,{\sc ii} may not be appropriate. Mg\,{\sc ii} transitions originate from the spontaneous decay of electrons to the ground state after excitation through either collisions or absorption of radiation. Specifically, the transitions occur from the $2P_{3/2}$ state for the 2796$\mathrm{\AA}$ line and the $2P_{1/2}$ state for the 2803$\mathrm{\AA}$ line. The Mg\,{\sc ii} line profile is complex due to emission originating from H II regions and the stars within them. Unlike Fe\,{\sc ii} lines, where the spectrum is relatively flat and the stellar continuum is primarily absorbed by the ISM, Mg\,{\sc ii} emission from H II regions contributes to the observed profile. Therefore, the nearby continuum is intrinsically different, as it is influenced by both local emission and propagation through the ISM and CGM.

            With these physical processes in mind, we modeled the Mg\,{\sc ii} doublet profile in three ways using Markov Chain Monte Carlo (MCMC) fitting \citep{2013Foreman}: first as a double Gaussian profile as pure absorption, then as a double Gaussian as pure emission, and last with four Gaussians, comprising simultaneously an emission and an absorption line component for each doublet member, as in \cite{Martin2013}. Our fitting routine normalized the flux and error to the absolute median, and then cropped the spectrum to the region surrounding the Mg\,{\sc ii} doublet. We then defined a likelihood function with centroid velocities, Gaussian widths, integrated fluxes, and the continuum level as parameters. 

            We constructed a prior function to constrain parameters based on the physical properties of the Mg\,{\sc ii} doublet. We set the absorption flux ratio constraint to $1 \le \mathrm{F}_{\mathrm{abs}}(\lambda2796)/\mathrm{F}_{\mathrm{abs}}(\lambda2803) \le 2$, with a ratio of 2 indicating optically thin gas and unsaturated lines, and a ratio of 1 corresponding to saturated lines. We allowed the emission flux ratio to vary from $0.8\le \mathrm{F}_{\mathrm{em}}(\lambda2796)/\mathrm{F}_{\mathrm{em}}(\lambda2803) \le 2.7$, since the dominant excitation mechanism determines the relative line intensity. Collisional excitation drives the ratio toward $\sim 2$, while radiative excitation brings the ratio closer to $\sim 1$, since both lines emit photons with equal probability. However, observed ratios frequently differ from theoretical values, suggesting that the intervening gas is composed of both uniform and porous gas that absorbs and scatters light \citep{Martin2012, Martin2013, Kornei2013, Chisholm2020}. We applied a prior linking the $\lambda2796$ and $\lambda2803$ wavelengths to preserve a consistent velocity structure across both transitions. To avoid non-physical values, we ensured that the absorption flux did not exceed the continuum minus the error. To prevent the widths from falling below the velocity dispersion of the galaxy, we required a minimum velocity width equal to or greater than H$\alpha$, or H$\beta$ when H$\alpha$ was unavailable.

            We then sampled the resulting posterior distribution using MCMC, and we report the median value of the samples for each parameter, with standard deviation of the sample distribution representing the uncertainty on the given parameter. We compared the $\chi^{2}$ among the three fits to identify the best fitting model. The two-component fit was used when the $p$-value was below 0.01, indicating a statistically significant improvement, justifying the inclusion of additional parameters. We determined the minimum flux threshold for including centroid measurements in our outflow analysis to be $8\sigma$. As for Fe\,{\sc ii}, we determined this threshold by testing a range of significance levels and selecting the lowest one greater than $3\sigma$ that yielded robust results upon actual inspection.

             Out of the 43 galaxies with Mg\,{\sc ii} coverage, we successfully fit Mg\,{\sc ii} profiles using MCMC to 16 of these galaxies (37\%), with 7 preferring the combination of emission and absorption fit and 9 preferring an absorption only fit. No galaxies show preference for the emission only fit, but a small number are still dominated by emission. The remaining 27 galaxies either lacked a detectable Mg\,{\sc ii} feature or exhibited an SNR that was too low for a robust fit. We found an average $\langle\Delta v_{\mathrm{Mg_{II}}}\rangle = -59\pm34$  km s$^{-1}$ using centroids from the absorption component.

        \subsubsection{Selection of Mg\,{\sc ii} Emitters} \label{sec:mgii_emission_methods}
            As shown in Figure \ref{fig:mgii_profiles}, the Mg\,{\sc ii} profile varies from absorption to emission dominated. Mg\,{\sc ii} emission results from photon scattering off the backside of galactic winds \citep{1993Phillips, 2009Weiner,2011Rubin}, similar to the Ly$\alpha$ emission \citep{2001Pettini, 2003Shapley, 2010Steidel, 2012Talia, 2012Erb, 2022Calabro, 2022Weldon, Kehoe2024}. 

            Various methods have been developed to identify galaxies with Mg\,{\sc ii} emission. \cite{2009Weiner} and \cite{2010Rubin} detected Mg\,{\sc ii} emission in continuum-normalized spectra at $z\sim1.4$ by measuring the flux above the continuum in two wavelength windows: one redwards of the $2803\mathrm{\AA}$ line and the other redwards of the $2796\mathrm{\AA}$ line. Their method yielded detection fractions of $\sim4\%$ \citep{2009Weiner} and $<1\%$ \citep{2010Rubin}, respectively. Other methods are more sensitive to weaker Mg\,{\sc ii} emission and have produced higher detection fractions. For example, \cite{2012Erb} examined star-forming galaxies at $1\le z \le2$. In their study,  Mg\,{\sc ii} emission was identified when at least two adjacent pixels exhibited flux at least $1.5\sigma$ above the continuum in either Mg\,{\sc ii} transition, leading to a detection rate of 30\%. This approach is highly sensitive to the S/N of the galaxy and the intrinsic strength of Mg\,{\sc ii} emission. \cite{Kornei2012} compared Fe\,{\sc ii} and Mg\,{\sc ii} profiles of $z\sim1$ star-forming galaxies to estimate the contribution of Mg\,{\sc ii} emission, reporting a detection fraction of $\sim15\%$. \cite{2018Feltre} used data from the MUSE \textit{Hubble} Ultra Deep Survey to identify Mg\,{\sc ii} emission at $0.7\le z \le 2.34$, selecting galaxies with an equivalent width of Mg\,{\sc ii} $<-1\,\mathrm{\AA}$ and $S/N>3$ in either Mg\,{\sc ii} component. \cite{2018Feltre} also distinguished between pure emission spectra and P-Cygni profiles, which combine emission and absorption, through visual inspection.

            For our analysis, we adopted the Mg\,{\sc ii} emission sample selected using the method from \cite{2009Weiner} and \cite{2010Rubin} as this method was the most conservative and selected the strongest Mg\,{\sc ii}  emitters in our data. We label this sample of 5 galaxies as ``Mg\,{\sc ii} emitters''.

    \subsection{Galaxy Properties}
        We derived various galaxy physical parameters, including $M_{*}$, dust attenuation, SFR, UV slope ($\beta$), specific star formation rate (sSFR = SFR/$M_{*}$) and $\Sigma_{\mathrm{SFR}}$ using a combination of spectral line measurements and photometry.

        Details on the stellar mass determinations are provided in \cite{Shapley2024}, with a brief summary given here. We estimated the stellar mass using SED modeling of photometry from \textit{JWST}/NIRCam, \textit{HST}/WFC3, and ACS, with catalogs from the DAWN \textit{JWST} Archive \citep{Heintz2025} and 3D-\textit{HST} \citep{2014Skelton}. SED modeling utilized the Fitting and Assessment of Synthetic Templates (FAST) program, incorporating the flexible stellar population synthesis models of \cite{2009Conroy} and a \cite{2003Chabrier} IMF. We assigned each galaxy to either a 1.4 solar metallicity model with the \cite{2000Calzetti} attenuation curve or a 0.27 solar metallicity model with the SMC extinction curve from \cite{2003Gordon}, labeled as ``$1.4Z_{\odot} +$ Calzetti'' and ``$0.27Z_{\odot} +$ SMC.'' For each galaxy, we selected the metallicity and extinction curves corresponding to the model with the lowest $\chi^{2}$ value.  We also corrected for the contributions of strong nebular emission lines \citep{2021Sanders} and nebular continuum emission \citep{2024Sanders}.

        We followed \cite{2024Clarke} to compute the SFR from the dust-corrected H$\alpha$ luminosity. In summary, to calculate the SFR, we first corrected for dust attenuation by assuming the \cite{1989Cardelli} dust law. We calculated $E(B-V)$ using the Balmer decrement, specifically the $\mathrm{H}\alpha/\mathrm{H}\beta$ ratio. We assumed an intrinsic $\mathrm{H}\alpha/\mathrm{H}\beta$ ratio of 2.79, consistent with Case B recombination at $T_{e}=15,000$ K. In cases where H$\alpha$ is not detected, we instead used the $\mathrm{H}\gamma/\mathrm{H}\beta$ ratio, assuming an intrinsic value of 0.475. For these galaxies, we first obtained the dust-corrected H$\beta$ flux and then inferred the corresponding H$\alpha$ flux using the intrinsic $\mathrm{H}\alpha/\mathrm{H}\beta$  ratio. If H$\beta$ was detected, we estimated the dust correction using the $\mathrm{H}\gamma/\mathrm{H}\alpha$ flux ratio, where the intrinsic value was 0.173. After correcting for dust attenuation, we converted H$\alpha$ luminosities to SFR using 
        \begin{equation}
            \log \left (\frac{\mathrm{SFR}}{M_{\odot}\mathrm{yr}^{-1}}\right) = \log \left (\frac{\mathrm{L}_{\mathrm{H}\alpha}}{\text{erg s}^{-1}}\right) + C
        \end{equation}
        with a metallicity-dependent conversion factor, $C$, determined from a set of BPASS models with an upper stellar mass limit of 100 $M_{\odot}$ \citep{2018Stanway, 2022Reddy}. As in \cite{2024Clarke}, we applied $C=-41.37$ for galaxies fit with the $1.4Z_{\odot} +$ Calzetti model and $C=-41.59$ for galaxies fit with the $0.27Z{\odot}$ + SMC model.  We calculated the UV continuum slope $\beta$ following the standard definition, $F_\lambda\propto\lambda^\beta$, by applying linear regression to the rest-frame UV photometry \citep{2024Clarke}. For each galaxy, we selected filters that probed the rest-frame wavelength range 1250--2600\,\AA. Uncertainties for $E(B-V)$, SFR, and $\beta$ were obtained from Monte Carlo distributions, where 10,000 random samples were drawn from the parameter distributions defined by the observed fluxes and their associated uncertainties.

        The $\Sigma_{\mathrm{SFR}}$ is expressed as:
        \begin{equation}
            \Sigma_{\text{SFR}} = \frac{\text{SFR}}{2\pi R_e^2}
        \end{equation}
        where $R_{e}$ represents the half-light effective radius. We determine $R_{e}$ by following the methodology of \cite{Pahl2022}, using \texttt{galfit} \citep{Peng2002,Peng2010} to model each galaxy's light profile with S\'ersic fits applied to image cutouts. For galaxies best described by a single S\'ersic component, we found the circularized effective radius as $R_e=r\sqrt{b/a}$, where $r$ is the semi-major axis half-light radius and $b/a$ is the axis ratio. For systems requiring multiple S\'ersic components, $R_e$ was derived from the area occupied by the brightest model pixels that account for half of the total flux with $A=\pi R_e^2$.
    
    \subsection{Composite Spectra}
        While the S/N for individual galaxies with line detections was 7.0 per pixel, the average S/N per pixel across the entire sample was 2.9. Therefore, we constructed composite spectra to increase the S/N and assess trends in outflow velocity from  Fe\,{\sc ii} or  Mg\,{\sc ii} centroids in relation to various stellar properties, which may be too weak to detect in individual measurements. We binned the spectra based on specific spectral properties, dividing them into ``low'' and ``high'' bins for each galaxy property: $M_{*}$, SFR, $\Sigma_{\mathrm{SFR}}$, sSFR, $\beta$, $E(B-V)$, and A$_{V}$. For each bin, we generated separate  Fe\,{\sc ii} and  Mg\,{\sc ii} stacks, where we included only spectra that cover the relevant spectral window (either around the  Fe\,{\sc ii} lines or the  Mg\,{\sc ii} doublet).  For the  Fe\,{\sc ii} stack, we focused exclusively on the $\lambda2587$ and $\lambda2600$ lines, as these exhibit the highest S/N. Additionally, we restricted our sample to galaxies with $z<4$ to ensure that the redshift distributions are consistent across bins. Each  Fe\,{\sc ii} bin contained 12 galaxies, while each  Mg\,{\sc ii} bin contained 13 galaxies. To create the composite spectra, we first shifted each individual, flux-calibrated galaxy spectrum into its rest frame. Then, we interpolated the spectra onto a common wavelength grid and scaled each spectrum so that its medium flux within a line-free rest frame window at 3100--3150\,\AA\, matched the mean scale of the bin. Lastly, computed the median flux at each wavelength for the entire bin.
        
        We measured the $\Delta v_{\mathrm{Fe_{II}}}$ and  $\Delta v_{\mathrm{Mg_{II}}}$ by fitting Gaussian curves (as described in Section \ref{sec:feii_fits}) and by MCMC fitting (as described in Section \ref{sec:mgii_fits}), respectively.

\section{Results}\label{sec:Results}
       Numerous studies have examined the outflow velocities of low-ionization interstellar absorption lines in low redshift galaxies for $z<3$ \citep{2005Rupke, 2009Weiner, 2010Chen, 2010Steidel, Kornei2012, 2013Bradshaw, 2015Chisholm, 2021Prusinski, 2022Weldon, Kehoe2024}. By extending to redshifts far beyond the scope of prior works, \textit{JWST} allows us to probe the evolution of Fe\,{\sc ii} and Mg\,{\sc ii} kinematics and compare the kinematics of Fe\,{\sc ii} and Mg\,{\sc ii} in a redshift range where these analyses were not previously possible.

    \subsection{Comparisons with Galaxy Properties}
         \begin{figure*}
            \centering            \includegraphics[width=\linewidth]{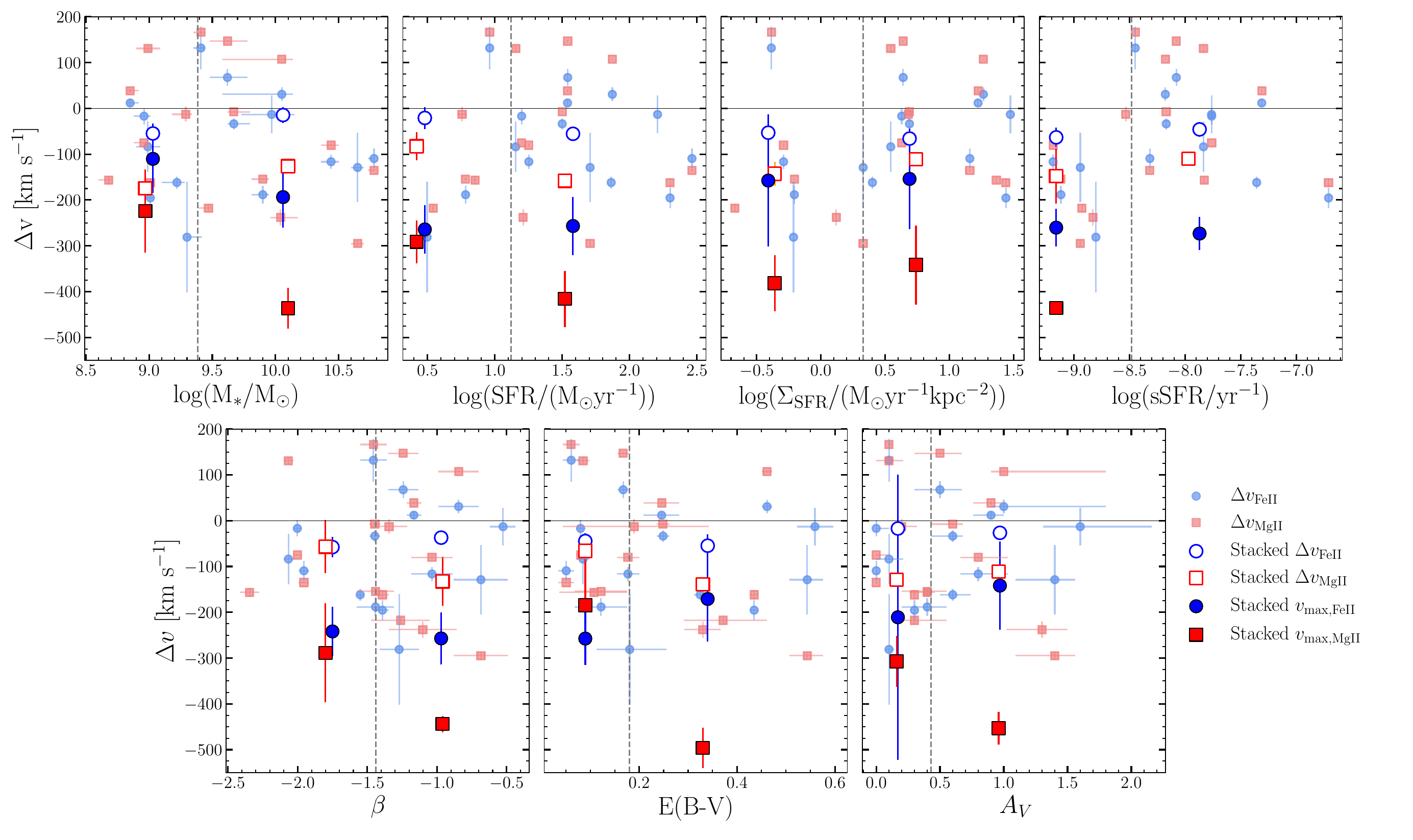}
            \caption{Outflow velocity versus various galaxy properties. Blue closed circles are velocities measured from Fe\,{\sc ii} centroids. Red closed squares are velocities from the Mg\,{\sc ii} doublet measured from MCMC. Large blue open circles are velocities from the Fe\,{\sc ii} stacks. Large red open squares are velocities from the Mg\,{\sc ii} stacks. Large red squares are $v_{\mathrm{max}}$ measured from the Mg\,{\sc ii} stacks. We were unable to robustly determine $v_{\mathrm{max}}$ for the high sSFR composite due to its complex profile shape. Large blue circles are $v_{\mathrm{max}}$ measured from the Fe\,{\sc ii} stacks. The dashed gray line indicates the separation between the low and high bins for each property. The individual measurements and composite spectra show no correlation between outflow velocity and galaxy property. The $v_\mathrm{max}$ measured from Mg\,{\sc ii}  suggests that the the amplitude of the outflow maximum velocity increases with $M_*$, SFR, $\beta$, $E(B-V)$, and $A_V$.} 
        \label{fig:scatter_vel}
    \end{figure*}
    
    \begin{deluxetable*}{lccccc}
    \tabletypesize{\small}
    \tablecaption{Correlation with Galaxy Properties\label{tab:p_values}}  
    \tablehead{
        \colhead{Galaxy Property} & \colhead{$\rho_{\mathrm{FeII}}$} & \colhead{$\rho_{\mathrm{MgII}}$} & \colhead{$D_{\mathrm{FeII}}$} & \colhead{$D_{\mathrm{MgII}}$}
    }
    \startdata
        $\log(M_{*})$ & $-0.043$ ($0.879$) & $-0.25$ ($0.35$) & $0.55$ ($0.02$) & $0.53$ ($0.02$) \\
        log(SFR) & $0.075$ ($0.719$) & $0.026$ ($0.922$) & $0.42$ ($0.14$) & $0.41$ ($0.12$)\\
        $\Sigma_{\mathrm{SFR}}$ & $0.23$ ($0.413$) & $0.31$ ($0.70$) & $0.31$ ($0.47$) & $0.23$ ($0.75$)\\
        sSFR & $0.17$ ($0.541$) & $0.33$ ($0.213$) & $0.25$ ($0.71$) & $0.36$ ($0.22$)\\
        $\beta$ & $0.10$ ($0.723$) & $-0.21$ ($0.444$) & $0.28$ ($0.59$) & $0.33$ ($0.31$)\\
        $E(B-V)$ & $-0.15$ ($0.594$) & $-0.42$ ($0.105$) & $0.22$ ($0.84$) & $0.23$ ($0.75$)\\
        A$_{V}$ & $0.16$ ($0.574$) & $-0.26$ ($0.337$) & $0.17$ ($0.98$) & $0.26$ ($0.61$) \\
    \enddata
    \tablecomments{The correlation coefficient $\rho$ (with $p$-values in parentheses) between $\Delta v_{\mathrm{FeII}}$ or $\Delta v_{\mathrm{MgII}}$ and various galaxy properties, and the Kolmogorov--Smirnov statistic $D$ (with $p$-values) comparing galaxies with detected outflows (from Fe\,{\sc ii} or Mg\,{\sc ii}) to those with no outflow detections.}
    \end{deluxetable*}
    To understand the driving mechanisms and evolution of galactic outflows, we investigate how outflow velocity relates to several galaxy properties such as $M_*$, SFR, sSFR, $\Sigma_\mathrm{SFR}$, $E(B-V)$, $\beta$, and A$_{V}$.
    
    The stellar mass represents the total mass of stars within a galaxy and serves as an  indicator of its evolutionary stage and gravitational potential. Massive galaxies represent deeper gravitational potential wells, which can influence the dynamics of outflows by regulating the escape of outflowing material \citep{2004Tremonti}. The SFR of a galaxy serves as a measurement of the amount of mechanical energy and radiation pressure available to drive galatic outflows. Combining the SFR and total stellar mass, the sSFR compares a galaxy's present star formation relative to its past average, offering a normalized measure of star formation activity. The $\Sigma_{\mathrm{SFR}}$ of a galaxy, which indicates the intensity of star formation per unit area, provides another key parameter in understanding outflow properties \citep{2000Heckman}. In regions with high $\Sigma_{\mathrm{SFR}}$, the radiation surface density is higher, making the radiation pressure on dust grains more effective in channeling energy and momentum from massive stars into the surrounding ISM to launch outflows.
    
    We also measured $\beta$, $E(B-V)$, and $A_{V}$, which serve as tracers of the galaxy's dust content. Specifically, $\beta$, where $F_\lambda \propto \lambda^{\beta}$, characterizes the slope of the UV continuum, which is sensitive to both dust attenuation and the age of the stellar population, $E(B-V)$ quantifies the reddening caused by dust absorption and scattering, and A$_{V}$ measures the total attenuation of starlight in the V-band by dust.

\begin{figure*}
        \centering
        \includegraphics[width=\linewidth]{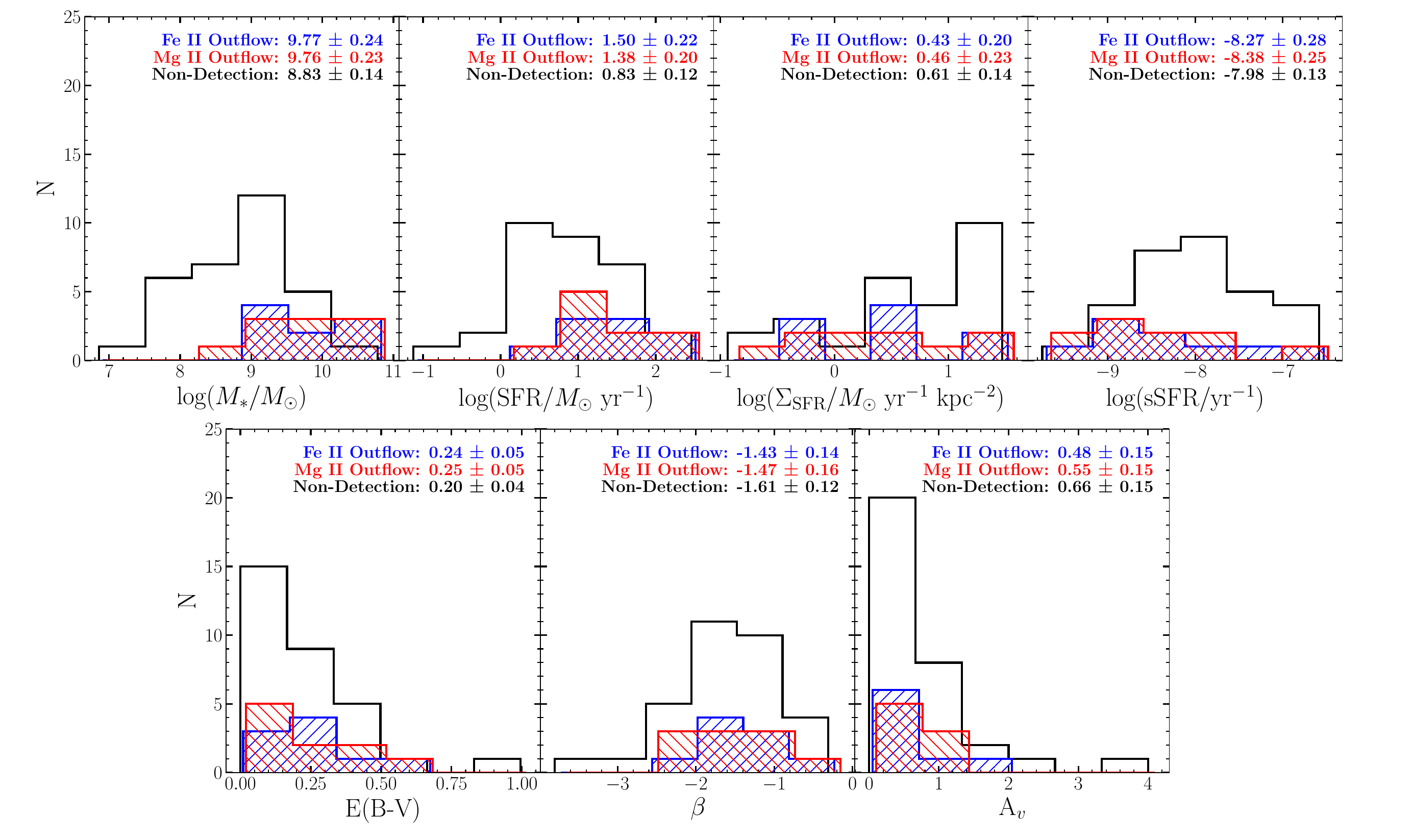}
        \caption{Comparisons of galaxies with detected outflows ($\Delta v <0$ and the magnitude of $\Delta v$ greater than its uncertainty) from Fe\,{\sc ii} lines (blue) or Mg\,{\sc ii} (red) and galaxies where no outflow was detected (black) for the various galaxy properties. Average property values are printed in the upper right corner. The average $M_*$ is significantly higher for galaxies with outflows detected from either Fe\,{\sc ii} or Mg\,{\sc ii}, compared with galaxies in which no outflow is detected.}
        \label{fig:outflow_detection_props}
    \end{figure*}

    In Figure \ref{fig:scatter_vel}, we compare $\Delta v_{\mathrm{Fe_{II}}}$ and $\Delta v_{\mathrm{Mg_{II}}}$ against the galaxy properties discussed above. We tested for correlations using the Spearman correlation coefficient, $\rho$ (Table \ref{tab:p_values}). For individual measurements, we did not find any significant correlations. To uncover any subtle trends, we compared the average properties of galaxies with outflows, identified by blueshifted absorption in either Fe\,{\sc ii} or Mg\,{\sc ii}, to those without any detected outflows (Figure \ref{fig:outflow_detection_props}). We applied the Kolmogorov–Smirnov (KS) test (Table \ref{tab:p_values}) and found that galaxies with detected outflows from either Fe\,{\sc ii} or Mg\,{\sc ii} show statistically significantly higher $M_{*}$\footnote{A more conservative test using the Bonferroni correction \citep{Simes1986}, which adjusts the significance threshold by dividing it by the number of comparisons to account for multiple tests, would deem this result insignificant.} (Figure \ref{fig:outflow_detection_props}). 

    \subsection{Mg\,{\sc ii} Emission}
    Studies analyzing galactic winds have extensively investigated  Mg\,{\sc ii} emission \citep{2009Weiner, 2010Rubin, 2011Rubin, Martin2012, 2012Erb, 2014Rigby, 2017Finley, 2018Feltre}. Mg\,{\sc ii} emission may originate from various processes, including active galactic nuclei (AGN) \citep{2009Weiner} or resonant scattering off the far side of the galaxy \citep{2009Weiner, 2012Erb}. Radiative transfer simulations of galactic-scale outflows from \cite{2011Prochaska} showed that blueshifted Mg\,{\sc ii} absorption was generally accompanied by emission near the systemic velocity of the galaxy. This emission partially filled in the absorption line profile, reducing the observed equivalent width and shifting the velocity closer to the systemic velocity of the galaxy. Their models predicted that significant Mg\,{\sc ii} emission was a generic feature of galactic winds, and that it could be observed under a wide range of conditions, including when dust, anisotropic flows, or limited observational apertures were present. These results suggest that the observed Mg\,{\sc ii} emission in star-forming galaxies mainly arises from resonantly scattered photons of outflowing gas.
    

    Detection rates for Mg\,{\sc ii} emission vary significantly across analyses for $1 \lesssim z \lesssim 2$, with most reporting fractions between $1\%$ and $30\%$ due to differences in selection methods. A detailed discussion of different selection methods is provided in Section \ref{sec:mgii_emission_methods}. Out of the 43 galaxies with  Mg\,{\sc ii}, we classify five as having significant Mg\,{\sc ii} emission, resulting in a detection fraction of $\sim12\%$. This is a much higher detection fraction than the $4\%$ and $<1\%$ reported by \cite{2009Weiner} and \cite{2010Rubin}, respectively, despite using the same method to identify Mg\,{\sc ii} emitters. The difference may stem from variations in sample selection, data quality, or differences in redshift and galaxy properties. However, since the detection fraction of Mg\,{\sc ii} emission is known to depend strongly on $M_*$, SFR, and rest-frame color, comparisons of detection fractions across surveys are only meaningful at fixed galaxy properties.

\begin{figure*}
        \centering
        \includegraphics[width=\linewidth]{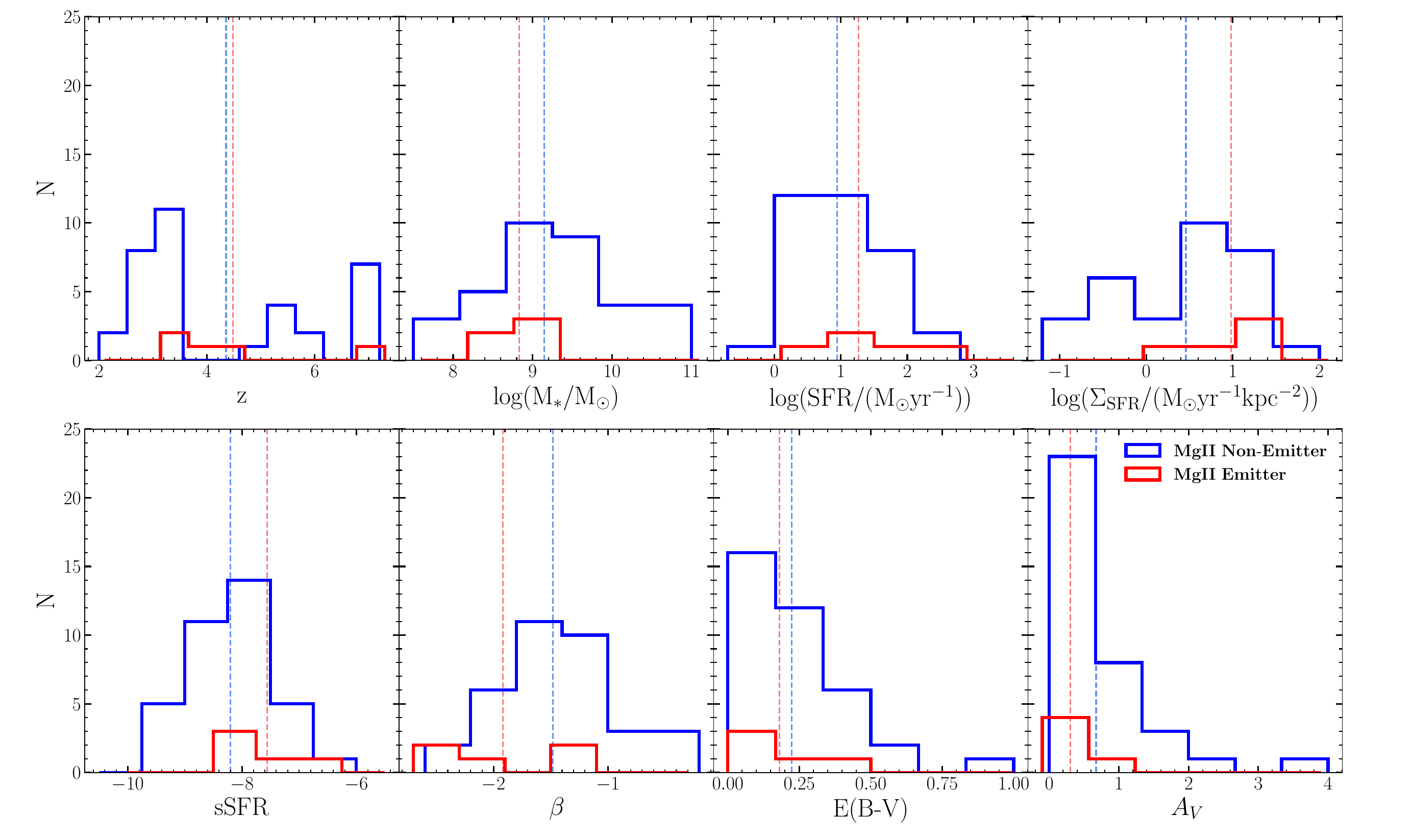}
        \caption{Histograms showing the distributions of various galaxy properties for the Mg\,{\sc ii} non-emitter (blue) and Mg\,{\sc ii} emitter (red) samples. The dashed blue and red lines show the property averages for non-emitters and emitters, respectively. On average, Mg\,{\sc ii} emitters have lower stellar mass, higher sSFR, lower $\beta$, and lower A$_{V}$ than Mg\,{\sc ii} non-emitters.}
        \label{fig:mgii_emitters}
    \end{figure*}

    In Figure \ref{fig:mgii_emitters}, we compare galaxies without significant Mg\,{\sc ii} emission (``Mg\,{\sc ii} non-emitters") to those in our Mg\,{\sc ii} emitter sample. We find that Mg\,{\sc ii} emitters have lower stellar masses, higher sSFR, lower $\beta$, and lower A$_{V}$ than Mg\,{\sc ii} non-emitters.

    \subsection{Trends with Composite Spectra}
    By analyzing individual galaxies, we found no significant trends between outflow velocity and any galaxy property. Additionally, we observed that galaxies exhibiting  Mg\,{\sc ii} emission tend to be bluer, have lower stellar masses, and have higher sSFRs compared to galaxies without  Mg\,{\sc ii} emission. We now investigate  the relationship between galaxy properties and the  Fe\,{\sc ii} and Mg\,{\sc ii} profiles by examining composite spectra binned by these properties.

        \subsubsection{Absorption Kinematics}
\begin{figure*}
            \centering
            \includegraphics[width=\linewidth]{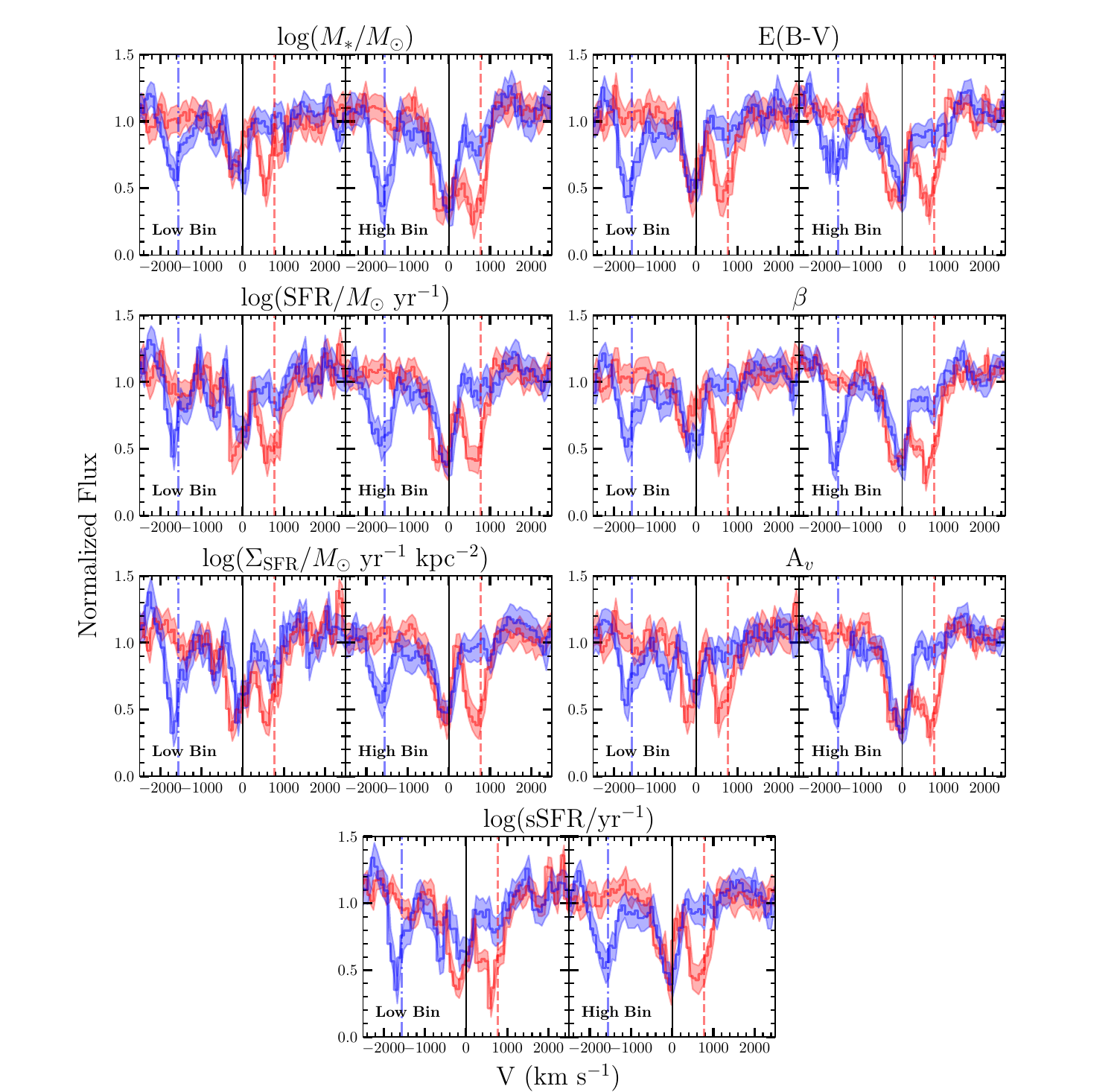}
            \caption{Comparisons of Fe\,{\sc ii} $\lambda2600$ (blue spectrum) and Mg\,{\sc ii} $\lambda2796$ (red spectrum) line profiles of the different composite spectra in velocity space. The solid vertical black line marks the systemic velocity, the dotted dashed blue line marks the location of Fe\,{\sc ii} $\lambda2587$ and the dashed red line marks the location of Mg\,{\sc ii} $\lambda2803$. In most spectra, there is an excess blue wing of the Mg\,{\sc ii}$\lambda2796$ profile compared to the Fe\,{\sc ii}$\lambda2600$ profile.} 
            \label{fig:composite_vel}
        \end{figure*}

        We investigate the Fe\,{\sc ii} and Mg\,{\sc ii} outflow velocities from the stacked spectra shown in Figures \ref{fig:scatter_vel} and \ref{fig:composite_vel}. As illustrated in Figure \ref{fig:scatter_vel}, 
        we find that the relative behavior of Fe\,{\sc ii} and Mg\,{\sc ii} across the two composite bins is generally consistent, but Mg\,{\sc ii} exhibits a systematically larger absolute offset, a trend that is also evident for individual galaxies that have both Fe\,{\sc ii} and Mg\,{\sc ii} measurements. When comparing the velocity profiles of Mg\,{\sc ii} and Fe\,{\sc ii} across different stacks (Figure \ref{fig:composite_vel}), we observe that Mg\,{\sc ii} often exhibits an excess blue wing relative to Fe\,{\sc ii}.

        To derive Mg\,{\sc ii} velocities from the composite spectra, we primarily use an absorption-only fit, except for the lower $\beta$ and A$_{V}$ bins, which preferred the combined absorption$+$emission model (Section \ref{sec:mgii_fits}). Since we rely on an absorption-only fit, the higher Mg\,{\sc ii} velocities may result from emission line filling. The Mg\,{\sc ii} $\lambda 2796$ absorption trough can be partially filled by $\lambda 2796$ emission, while the $\lambda 2803$ absorption trough can receive contributions from both transitions. Specifically, the red wing of $\lambda 2796$ emission can fill the blue wing of the $\lambda 2803$ absorption trough, while $\lambda 2803$ emission can fill the red wing of the absorption trough.

        Emission line filling occurs because Mg\,{\sc ii} emission is often redshifted due to photon scattering off the far side of galactic winds, whereas Mg\,{\sc ii} absorption is typically blueshifted \citep{Kornei2013, Martin2013}. Since Mg\,{\sc ii} transitions are resonantly trapped, photons are repeatedly re-emitted without alternative decay paths, leading to emission filling \citep{Kornei2012}. In contrast to Mg\,{\sc ii} transitions, Fe\,{\sc ii} lines can either re-emit a photon at the same wavelength as the absorbed photon or decay to intermediate fine-structure states (Fe\,{\sc ii}* emission). The existence of these fine-structure levels prevents the absorption lines from being filled with scattered, re-emitted photons, thus avoiding emission line filling \citep{2011Prochaska}. Emission line filling of Mg\,{\sc ii} has been observed in multiple research works. For instance, \cite{Martin2012} found that Mg\,{\sc ii} outflow velocities are significantly higher than Fe\,{\sc ii} and frequently display a blue wing, which they attribute to emission filling. Similarly, \cite{2012Erb} reported evidence of emission filling in both individual spectra and composite spectra binned by galaxy properties. Their study found that galaxies with detectable Mg\,{\sc ii} emission exhibited an average velocity shift of $-337 \pm 17$ km s$^{-1}$, while those without detectable emission had a shift of $-189 \pm 24$ km s$^{-1}$. In our sample, we find that Mg\,{\sc ii} emitters have an average velocity shift of $-180 \pm 30$ km s$^{-1}$, whereas non-emitters exhibit a shift of $-77 \pm 12$ km s$^{-1}$.
        
        Due to the potential for emission line filling in Mg\,{\sc ii}, several studies use the maximum velocity ($v_{\mathrm{max}}$) of the $\lambda2796$ line to study Mg\,{\sc ii} \citep{Martin2012, 2012Erb, Kornei2013}. To determine the $v_{\mathrm{max}}$ of Mg\,{\sc ii} we followed the method described by \cite{Kornei2012}. In summary, we first identified the minimum of the $2796\mathrm{\AA}$ absorption trough. Moving toward shorter wavelengths, we evaluated the sum of the flux and the uncertainty at each step until the sum exceeded a threshold set by the continuum S/N, defined as $1.0-1.0/\mathrm{(S/N)}$. $v_{\mathrm{max}}$ was determined by the wavelength at which this condition was first met. To account for uncertainties, we applied a Monte Carlo approach, perturbing the spectrum 1000 times by adding noise drawn from a Gaussian distribution with a standard deviation equal to the flux uncertainty. We find that Mg\,{\sc ii} emitters have an average $v_{\mathrm{max}}$ of $-643\pm 108$ km s$^{-1}$, while Mg\,{\sc ii} non-emitters have an average $v_{\mathrm{max}}$ of $-353 \pm 25$ km s$^{-1}$. We apply the same methodology to Fe\,{\sc ii}$\lambda2587 \text{ and }\lambda2600$ and take the inverse-variance-weighted average of the two lines. For Fe\,{\sc ii} we find an average $v_{\text{max}}$ of $-307\pm35\,\text{km s}^{-1}$.

        \subsubsection{Trends with Galaxy Properties}
        \begin{deluxetable*}{lccccc}
        \tablecaption{Composite Spectra Measurements}
        \label{tab:comp}
        \tablehead{
        \colhead{Composite} & \colhead{Average Property Value} & \colhead{$\Delta v_{\text{FeII}}$} & \colhead{$\Delta v_{\text{MgII}}$} & \colhead{$v_{\text{max,FeII}}$} & \colhead{$v_{\text{max,MgII}}$} \\
         & \colhead{and Range} & \colhead{$\left[\text{km\,s}^{-1}\right]$} & \colhead{$\left[\text{km\,s}^{-1}\right]$} & \colhead{$\left[\text{km\,s}^{-1}\right]$} & \colhead{$\left[\text{km\,s}^{-1}\right]$}
        } 
        
        \startdata
            Low $\log(M_*/M_\odot)$ & $8.97\,(8.65-9.36)$ & $-55\pm21$ & $-174\pm16$ & $-110\pm74$ & $-224\pm91$ \\
            High $\log(M_*/M_\odot)$ & $10.10\,(9.41-10.78)$ & $-14\pm17$ & $-126\pm11$ & $-194\pm66$ & $-436\pm45$ \\
            Low $\log(\text{SFR}/(M_\odot \text{yr}^{-1}))$ & $0.42\,((-0.10)-0.96)$ & $-21\pm24$ & $-83\pm30$ & $-264\pm53$ & $-291\pm47$ \\
            High $\log(\text{SFR}/(M_\odot \text{yr}^{-1}))$ & $1.52\,(1.16-2.46)$ & $-55\pm16$ & $-158\pm10$ & $-257\pm63$ & $-416\pm61$ \\
            Low $\log(\Sigma_{\mathrm{SFR}}/(M_\odot \text{yr}^{-1}\text{kpc}^{-2}))$ & $-0.36\,((-1.55)-0.32)$ & $-53\pm18$ & $-143\pm27$ & $-158\pm144$ & $-382\pm61$ \\
            High $\log(\Sigma_{\mathrm{SFR}}/(M_\odot \text{yr}^{-1}\text{kpc}^{-2}))$ & $0.74\,(0.34-1.26)$ & $-66\pm21$ & $-111\pm10$ & $-154\pm110$ & $-342\pm86$ \\
            Low $\log(\text{sSFR}/\text{yr}^{-1})$ & $-9.16\,((-11.00)-(-8.59)))$ & $-63\pm21$ & $-148\pm61$ & $-260\pm42$ & $-436\pm9$ \\
            High $\log(\text{sSFR}/\text{yr}^{-1})$ & $-7.97\,((-8.45)-(-7.23))$ & $-46\pm16$ & $-110\pm12$ & $-273\pm37$ & $...$ \\
            Low E(B-V) & $0.09\,(0.00-0.17)$ & $-45\pm17$ & $-66\pm12$ & $-258\pm58$ & $-185\pm100$ \\
            High E(B-V) & $0.33\,(0.18-0.54)$ & $-55\pm24$ & $-139\pm14$ & $-171\pm93$ & $-496\pm44$ \\
            Low $\beta$ & $-1.80\,((-2.15)-(-1.45))$ & $-58\pm22$ & $-57\pm58$ & $-242\pm54$ & $-289\pm108$ \\
            High $\beta$ & $-0.96\,((-1.44)-0.07)$ & $-37\pm14$ & $-133\pm54$ & $-257\pm57$ & $-444\pm18$ \\
            Low $A_{V}$ & $0.16\,(0.00-0.40)$ & $-17\pm46$ & $-129\pm12$ & $-211\pm311$ & $-307\pm56$ \\
            High $A_{V}$ & $0.96\,(0.40-1.80)$ & $-27\pm13$ & $-111\pm9$ & $-142\pm96$ & $-453\pm35$ \\
        \enddata
        \tablecomments{We were unable to robustly determine $v_{\mathrm{max}}$ for the high sSFR composite due to its complex profile shape.}
        \end{deluxetable*}

        Composite spectra offer valuable insights into how various galaxy properties, such as $M_*$, SFR, $\Sigma_{\mathrm{SFR}}$, sSFR, $\beta$, $E(B-V)$, and A$_{V}$, influence outflow velocities and line profiles that may be hidden in individual spectra.  In Figure \ref{fig:scatter_vel}, we show velocities of Fe\,{\sc ii} (blue open circles) and Mg\,{\sc ii} (red open squares) and $v_{\text{max}}$ of Fe\,{\sc ii} (blue circles) and Mg\,{\sc ii} (red squares) measured from the composite spectra. Specific velocities for each composite are listed in Table \ref{tab:comp}. We observe potential weak trends in $v_{\text{max}}$ indicating that higher $M_*$, SFR, $\beta$, and $E(B-V)$, and $A_V$ may correspond to more blueshifted $v_{\text{max}}$.

\begin{figure*}
            \centering
            \includegraphics[width=\linewidth]{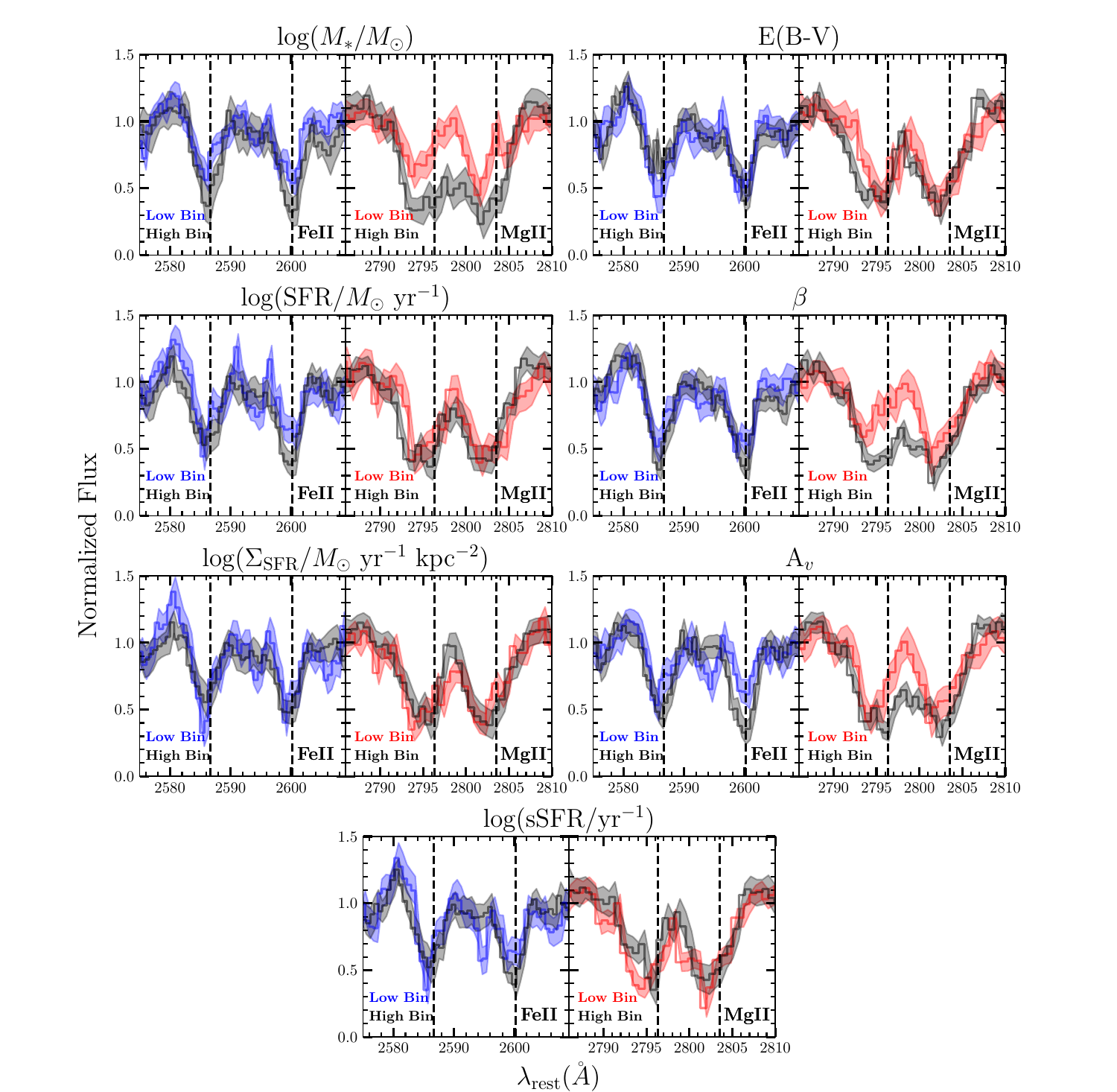}
            \caption{Comparisons of the low (blue for the Fe\,{\sc ii} composite and red for the Mg\,{\sc ii}) and high (black) bins for the different stacks. The left plot on the subplots are the Fe\,{\sc ii} stacks and the right plots on the subplots are the Mg\,{\sc ii} stacks. The dashed vertical lines mark the rest wavelengths for Fe\,{\sc ii} $\lambda2587$, Fe\,{\sc ii}$\lambda2600$, Mg\,{\sc ii} $\lambda2796$ and Mg\,{\sc ii} $\lambda 2803$, respectively. The Mg\,{\sc ii} stacks for $M_*$, $\beta$, and A$_{V}$ exhibit the strongest evidence of Mg\,{\sc ii} emission in the low bins.}
            \label{fig:composite_profiles}
        \end{figure*}

\begin{figure*}
            \centering
            \includegraphics[width=\linewidth]{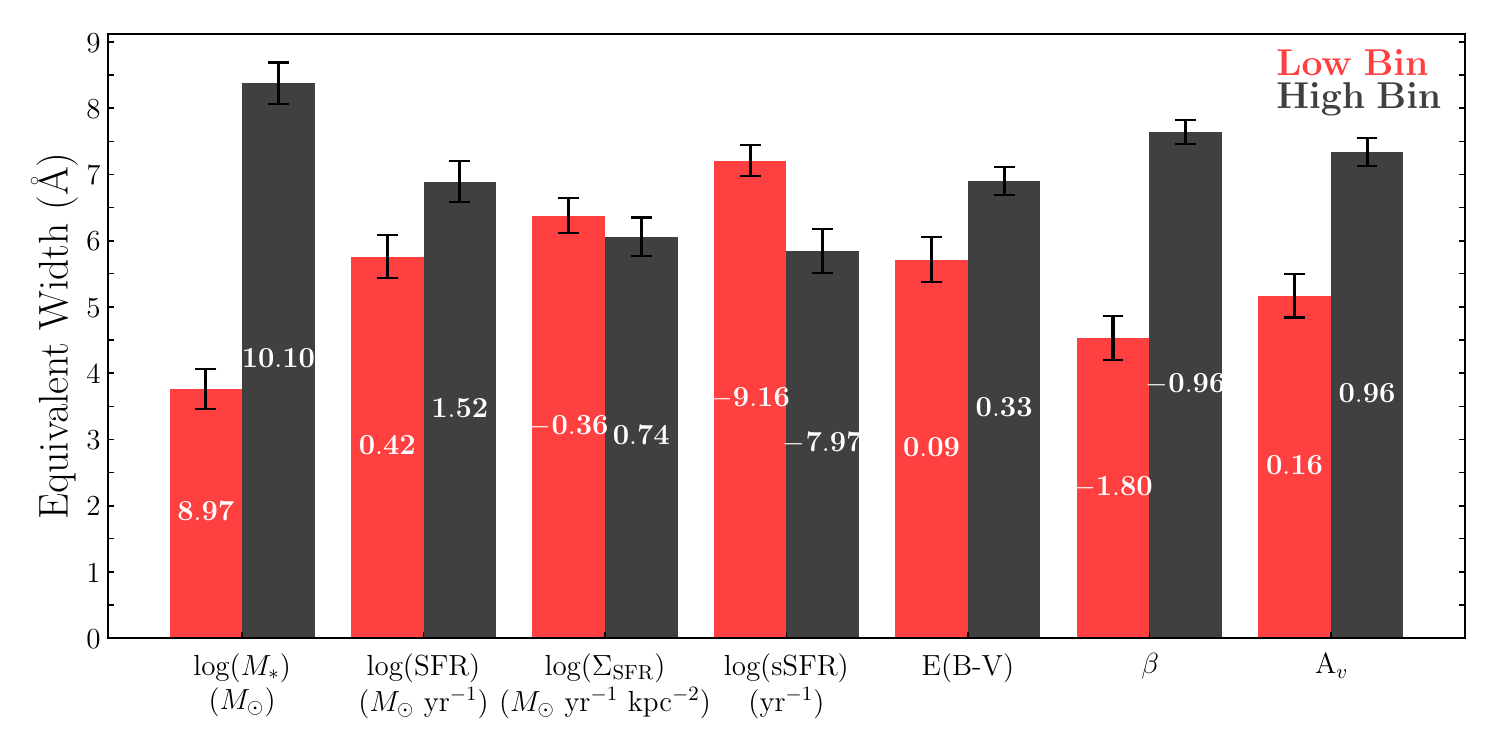}
            \caption{Equivalent width (EW) comparisons of the Mg\,{\sc ii} low bin (red) and high bin (black) stacks. The white numbers inside each bar are the average property values of each stack. The high bins of the stellar mass, $\beta$, and A$_{V}$ have much greater EWs than the low bins, whereas sSFR shows the inverse relationship.}
            \label{fig:mgii_EW}
        \end{figure*}

        We also use composite spectra to identify trends between Mg\,{\sc ii} emitters and non-emitters. In Figure \ref{fig:composite_profiles}, we compare the Fe\,{\sc ii} ($\lambda2587$ and $\lambda2600$) and Mg\,{\sc ii} line profiles, where the low property bin is represented in blue (Fe\,{\sc ii}) and red (Mg\,{\sc ii}), while the high property bin is shown in black. For Fe\,{\sc ii}, we do not observe any significant trends between the low and high bins for any property. In contrast, for Mg\,{\sc ii}, we detect evidence of emission in the low bins for stellar mass, $\beta$, and $E(B-V)$. As shown in Figure \ref{fig:mgii_EW}, the EW of Mg\,{\sc ii} is significantly greater in the high bins for stellar mass, $\beta$, and A$_{V}$.

\section{Discussion}\label{sec:Discussion}
    \subsection{Outflow Velocity and Galaxy Property Correlations}

      We compare our results to established trends between outflow velocities and galaxy properties derived from Fe\,{\sc ii} and Mg\,{\sc ii} at lower redshifts, as well as trends derived from FUV lines at higher redshifts more comparable to those spanned by our sample. These comparisons help contextualize our findings in terms of the dominant outflow driving mechanisms and highlight how limited dynamic range and sample size may obscure expected correlations.
    
    \subsubsection{Correlations with Stellar Mass}
    We find no significant correlation between outflow velocity and stellar mass based on NUV absorption lines in both individual galaxies and composite spectra. This is consistent with previous results at similar redshifts \citep{2010Steidel,2012Talia,2022Calabro,Kehoe2024}. However, galaxies with detected Fe\,{\sc ii} and Mg\,{\sc ii} outflows tend to have higher stellar masses compare to those without an outflow detection (Figure \ref{fig:mgii_emitters} and Table \ref{tab:p_values}). Additionally, Mg\,{\sc ii} composite spectra show a significant increase in $v_{\text{max}}$ with stellar mass (Figure \ref{fig:scatter_vel}), in agreement with trends reported by \citet{2012Erb}.

    Outflow velocity is influenced by both gravitational potential depth and the energy injected by star formation and feedback. More massive galaxies typically host higher star formation rates, which can generate stronger feedback capable of driving faster outflows despite deeper potential wells. \citet{2009Weiner} and \citet{2012Erb} found that $v_{\text{max}}$ increases with stellar mass, with $v_{\text{max}} \propto M_*^{0.17}$. Within the uncertainties, our $v_{\text{max}}$ results are consistent with this relation. This suggests that the stronger energy and momentum input from star formation and associated feedback processes in massive galaxies can overcome their deeper gravitational potential wells, resulting in higher outflow velocities.
    
    At $z \sim 1$, several studies using NUV absorption lines have reported positive correlations between stellar mass and outflow velocity \citep{2009Weiner,2010Rubin,2012Erb}. For example, \cite{2010Rubin} observed stronger Mg\,{\sc ii} blueshifts in more massive galaxies using a sample of 468 galaxies between $0.7 < z < 1.5$. However, other low-redshift studies have not found significant correlations \citep{Martin2012,2014Bordoloi,2021Prusinski}, possibly due to differences in sample selection, spectral resolution, or contamination from stellar absorption.
    
    At higher redshifts ($z > 1$), previous analyses of FUV lines typically found no significant correlation between outflow velocity and stellar mass. \citet{2012Talia} studied 74 star-forming galaxies from the GMASS survey at $z \sim 2$ and found no clear trend. \cite{2022Calabro} reached a similar conclusion using 330 galaxies from the VANDELS survey at $2 \leq z \leq 5$, both in individual and stacked spectra. The lack of observed correlations in these studies, as well as in our individual measurements, may arise from limited sample sizes and relatively large measurement uncertainties. Expanding to larger, higher signal-to-noise samples will be essential for analyzing the relationship between outflow velocity and stellar mass.
   
     \subsubsection{Correlations with SFR}
     We find no significant correlation between outflow velocity and SFR, and no evidence that galaxies with detected outflows have higher average SFRs than those without (Figure \ref{fig:outflow_detection_props} and Table \ref{tab:p_values}). The lack of trend between outflow velocity and SFR is likely due to the limited dynamic range in SFR within our sample, ($10-100M_\odot\,\mathrm{yr}^{-1}$). This range is consistent with those in previous high-redshift studies that similarly report weak or no correlations between outflow properties and SFR \citep{2010Steidel,2022Calabro}. Furthermore, in the local universe at $z\sim0.1$, \cite{2005Martin} demonstrated that a clear correlation between outflow velocity and SFR emerged only when they extended their sample to include dwarf galaxies with SFR $<1M_\odot\,\text{yr}^{-1}$.

     Although no trends are observed in the measurements and distributions of individual galaxies, our composite spectra reveal that outflow velocities traced by both Fe\,{\sc ii} and Mg\,{\sc ii} increase with SFR. Additionally, $v_{\text{max}}$ derived from the composite Mg\,{\sc ii}  line profiles also exhibits an increase with SFR (Figure \ref{fig:scatter_vel} and Table \ref{tab:comp}). The scaling relationship between outflow velocity and SFR depends on the dominant driving mechanism of the outflow. Specifically, outflows driven by mechanical energy injected by overlapping supernovae are predicted to produce outflow velocities that scale weakly with SFR (approximately $v_{\text{out}}\propto\text{SFR}^{0.2-0.25}$)\citep{2000Heckman,2006Ferrara}. In contrast, momentum-driven outflows, dominated by radiation pressure on dust grains and momentum injection from supernovae, are expected to show a stronger, approximately linear scaling with SFR \citep{2012Sharma}. The weak dependence in our composite spectra suggests that the outflows are primarily energy-driven, where our velocities are consistent with the energy-driven scaling relation given our measurement uncertainties.

     Observational studies using NUV tracers at $z \sim 1$ have generally found faster outflows in galaxies with higher SFRs, particularly in stacked spectra \citep{2009Weiner,2010Rubin,Martin2012,2013Bradshaw,2014Rubin,2014Bordoloi,2021Prusinski}. Specifically, \cite{Martin2012} split a sample of 200 galaxies at $0.4\leq z \leq 1.4$ into thirds and found that higher outflow velocities were more prevalent among galaxies with the highest SFRs compared to those with the lowest SFRs. Similarly, \cite{2021Prusinski} studied 22 galaxies between $1\lesssim z\lesssim1.5$ and reported a velocity difference of $\Delta v = -77 \pm 39\,\text{km\,s}^{-1}$ between high- and low-SFR composites.

    At higher redshifts ($z \sim 2$), FUV studies similarly report weak positive trends between outflow velocity and SFR. \cite{2022Weldon} measured $v_{\mathrm{max}}$ in 155 star-forming galaxies and found a weak correlation with SFR, while \cite{Kehoe2024} observed significantly higher average SFRs among galaxies with detected outflows in a sample of 80 galaxies at $1.4 \leq z \leq 2.7$. In contrast, several other studies using FUV tracers report no significant correlation, most likely due to the limited dynamic ranges in SFR of their samples \citep{2010Steidel,2012Talia,2022Calabro}. 

    These findings suggest that weak trends between outflow velocity and SFR are common in both NUV tracers at $z \sim 1$ and FUV tracers at $z \sim 2$, particularly when the data spans a sufficiently broad range in SFR. Our results are consistent with previous works, where detectable trends between outflow velocity and SFR require either large, diverse samples or composite spectra analyses, and are most consistent with energy-driven outflow scaling relations.
    
     \subsubsection{Correlations with sSFR}
     We find no strong correlation between outflow velocity and sSFR when examining the individual velocity measurements and distributions (Figure \ref{fig:scatter_vel}, \ref{fig:outflow_detection_props} and Table \ref{tab:p_values}). Furthermore, the velocities are consistent among the high- and low-sSFR composite bins (Figure \ref{fig:scatter_vel} and Table~\ref{tab:comp}).

     Given that sSFR reflects both the mechanical energy available from star formation and the gravitational potential depth associated with stellar mass, a correlation with outflow velocity is physically motivated. At $z\sim1$, several analyses using NUV tracers have found trends between outflow velocity and sSFR. For example, \cite{2013Bradshaw}, analyzed composite spectra of 413 $K$-band selected galaxies a $0.71\leq z \leq 1.63$, finding that galaxies in the highest-sSFR bin exhibited significantly faster outflows than those with lower sSFRs. Similarly, \cite{Martin2012} showed that galaxies with $-8.8 < \log(\text{sSFR/yr}^{-1})<-8.2$ were about five times more likely to display blueshifted Fe\,{\sc ii} absorption than those with $-10< \log(\text{sSFR/yr}^{-1})<-9$, indicating a strong link between star formation intensity and the prevalence of winds.

     Other studies using NUV lines have reported a lack of correlation between sSFR and outflow velocity \citep{2010Rubin, 2021Prusinski}. Similarly, in the FUV at $z\sim2$, \cite{2022Weldon} found no significant relationship between sSFR. Our results therefore align with some previously reported in the literature, in which there is yet to be a clear consensus. Higher-quality data or a larger dynamic range may be necessary to uncover any subtle dependencies between sSFR and outflow velocity.
   
     \subsubsection{Correlations with $\Sigma_{\text{SFR}}$}
         When analyzing the relationship between outflow velocity and $\Sigma_{\text{SFR}}$, we do not observe a significant trend with individual velocity measurements or the distributions of $\Sigma_{\text{SFR}}$ between galaxies with and without detected outflows (Figure \ref{fig:scatter_vel}, \ref{fig:outflow_detection_props} and Table \ref{tab:comp}). Velocities measured from the composite spectra are also consistent across the low- and high-$\Sigma_{\text{SFR}}$ bins (Table \ref{tab:comp}). This lack of correlation is likely due to the limited dynamic range of $\Sigma_{\text{SFR}}$ in our sample. Outflows are generally thought to require a minimum $\Sigma_{\text{SFR}}$ threshold of $\sim 0.1\,M_\odot\text{yr}^{-1}\text{kpc}^{-2}$ to be launched \citep{2002Heckman}.

    Similar to SFR, there are different scaling relations between outflow velocity and $\Sigma_{\text{SFR}}$, depending on the dominant driving mechanism of the outflow. Energy-driven models predict a relatively weak dependence of outflow velocity on $\Sigma_{\text{SFR}}$, scaling as $v_{\text{out}} \propto \Sigma_{\text{SFR}}^{0.1}$ \citep{2010Chen}. In contrast, momentum-driven winds produce much steeper relations, such as $v_{\text{out}} \propto \Sigma_{\text{SFR}}^{2}$ \citep{2011Murray}. The scaling may also fall somewhere in between,  reflecting a mixture of driving mechanisms. Our observed lack of a strong trend between outflow velocity and $\Sigma_{\text{SFR}}$ suggests that the winds in our sample may be energy-driven.

   Kinematic studies using Fe\,{\sc ii} and Mg\,{\sc ii} absorption at $z\sim1$ have found that outflow velocities increase with higher $\Sigma_\mathrm{SFR}$ \citep{2010Rubin, Kornei2012, 2014Bordoloi, 2021Prusinski}. \cite{Kornei2012} analyzed 72 galaxies in the Extended Groth Strip and found the strongest correlation between outflow velocity and $\Sigma_\mathrm{SFR}$, with a significant $3.1\sigma$ trend. Similarly, \cite{2014Bordoloi} used Mg\,{\sc ii} absorption in a sample of 486 galaxies at $1 \leq z \leq 1.5$ and reported higher outflow velocities in galaxies with higher $\Sigma_\mathrm{SFR}$.

   FUV absorption and H$\alpha$ emission studies also support this trend \citep{2019Davies, 2019Schreiber, 2022Calabro, Kehoe2024}. \cite{2019Davies} derived a power-law relationship of $v_{\text{out}} \propto \Sigma_{\text{SFR}}^{0.34}$ between H$\alpha$ outflow velocity and $\Sigma_\mathrm{SFR}$ in 28 galaxies at $2 \lesssim z \lesssim 2.6$, while \cite{2019Schreiber} observed more frequent outflows in galaxies with higher $\Sigma_\mathrm{SFR}$ in a larger sample of $\sim$600 galaxies at $0.6 < z < 2.7$. Since H$\alpha$ emission traces denser, ionized gas than the FUV/NUV absorption lines, it may be expected that outflows measured from H$\alpha$ emission are more directly associated with the current SFR.

    Similar to our results, some FUV-based studies do not find a significant correlation \citep{2010Steidel, 2022Weldon}, likely due to limited dynamic range in $\Sigma_\mathrm{SFR}$. As in our sample, both \cite{2010Steidel} and \cite{2022Weldon} lack coverage below the \cite{2002Heckman} $\Sigma_{\text{SFR}}$ threshold. A larger dynamic range may be necessary to probe the relations between outflow velocity and $\Sigma_{\text{SFR}}$.
     
    \subsection{Comparing Mg\,{\sc ii} Emitters and Non-Emitters}
        When comparing our sample of Mg\,{\sc ii} emitters at $z\gtrsim2$ and non-emitters in Figure \ref{fig:mgii_emitters}, we observe that Mg\,{\sc ii} emitters have lower $M_*$, higher sSFRs, and possess bluer (more negative $\beta$ and less attenuated (lower A$_V$)) stellar populations than non-emitters. Furthermore, our composite spectra show similar trends (Figure \ref{fig:composite_profiles}). We find that composites of galaxies with higher $M_*$, $\beta$, and $A_V$ exhibit significantly larger Mg\,{\sc ii} absorption EWs compared to those with lower values of these properties (Figure \ref{fig:mgii_EW}). We find that these correlations persist at $z\gtrsim2$, extending well-established trends from lower redshifts at $z\lesssim2$ to earlier epochs \citep{2010Rubin, 2012Erb, Martin2012, Martin2013, Kornei2012, Kornei2013, 2018Feltre}. This suggests that Mg\,{\sc ii} emission remains a reliable tracer of low-mass, actively star-forming, and relatively dust-poor galaxies across a broad redshift range. Moreover, our results also indicate that Mg\,{\sc ii} emission is sensitive to ISM properties such as dust and the neutral gas distribution.
        
    These observed correlations suggest that the Mg\,{\sc ii} emitters in our sample may serve as promising sources of leaking Lyman Continuum (LyC) emission. Since Mg\,{\sc ii} emission is a resonant transition with an ionization potential of 15 eV, which closely matches that of neutral hydrogen, it is a useful indirect tracer of the Lyman Continuum escape fraction ($f_{\text{esc}}$). Recent work has explored using Mg\,{\sc ii} as an indirect tracer of $f_{\text{esc}}$ \citep{2018Henry,Chisholm2020,2022Katz}. \cite{Chisholm2020} analyzed 10 LyC emitters in the local universe at $z\lesssim0.50$ to investigate whether the Mg\,{\sc ii} doublet could serve as an estimator for $f_{\text{esc}}$. They found that the combination of Mg\,{\sc ii} emission strength and FUV dust attenuation accurately estimated $f_{\text{esc}}$. In high-resolution simulations of $z=6$ star-forming galaxies, \cite{2022Katz} found that galaxies with significant LyC leakage also tended to have Mg\,{\sc ii} observed in emission. However, the interpretation of the Mg\,{\sc ii} doublet ratio in terms of $f_{\text{esc}}$ is not straightforward, due to the interaction between the dust, gas and more realistic ISM geometries.
    
    \subsection{Comparing Outflow tracers}
    
    \subsubsection{Near-UV vs. Far-UV Tracers of ISM Gas Kinematics}
     We identify one galaxy in our sample at $z=5.19$, GOODSN-100067, for which both NUV and FUV line centroid measurements are available. In the FUV, we detect Al\,{\sc ii} $\lambda1670$, while in the NUV, we measure Fe\,{\sc ii} $\lambda$, 2344, 2383, 2587, and 2600. For this galaxy, the measured line centroids are redshifted relative to the systemic velocity, suggesting the presence of inflowing rather than outflowing gas. Specifically, we find that the FUV velocity is $+43.1 \pm 25.3$ km s$^{-1}$, and the NUV velocity is $+8.1 \pm 8.8$ km s$^{-1}$. FUV and NUV transitions yield kinematic results that are consistent within $2\sigma$, suggesting that NUV and FUV tracers can yield comparable measurements of ISM kinematics. Expanding this subset of galaxies with both NUV and FUV measurements will facilitate a more detailed comparison of the kinematics traced by different wavelength regimes.
    
    Historically, galactic outflows have been studied primarily using NUV absorption lines (e.g., Fe\,{\sc ii} and Mg\,{\sc ii}) for galaxies at redshifts $z\sim0.5-2$ \citep{2009Weiner, 2010Rubin, 2011Prochaska, 2012Erb, Kornei2012, Martin2012, Martin2013, Kornei2013, 2014Bordoloi, 2014Rubin, 2018Feltre,  2021Prusinski}, and FUV absorption lines (e.g., low-ionization transitions including Si\,{\sc ii}, O\,{\sc i}, C\,{\sc ii}, O\,{\sc i}, Al\,{\sc ii}, and high-ionization features such as Si\,{\sc iv} and C\,{\sc iv}) for galaxies at $z>2$ \citep{2003Shapley, 2010Steidel, 2012Talia, 2019Davies, 2022Calabro, 2022Weldon, Kehoe2024}. The limited wavelength coverage over which ground-based telescopes obtain high S/N detections of continuum emission has made it challenging to analyze NUV absorption features for galaxies beyond $z\sim 2$.
    
    \textit{JWST} overcomes this limitation, enabling the study of NUV features in high-redshift galaxies, which allows for direct and uniform comparisons to lower-redshift work using the same tracer of ISM kinematics. In addition to extending NUV analyses to much higher redshifts, the broad spectral coverage of \textit{JWST} allows for simultaneous observations of both NUV and FUV features within a single spectrum. This capability allows for a direct comparison of NUV and FUV diagnostics in individual high-redshift galaxy spectra for the first time, providing insight into whether these probes yield consistent measurements or reveal systematic differences.

    \subsubsection{Na\,D}
\begin{figure}
        \centering
        \includegraphics[width=\linewidth]{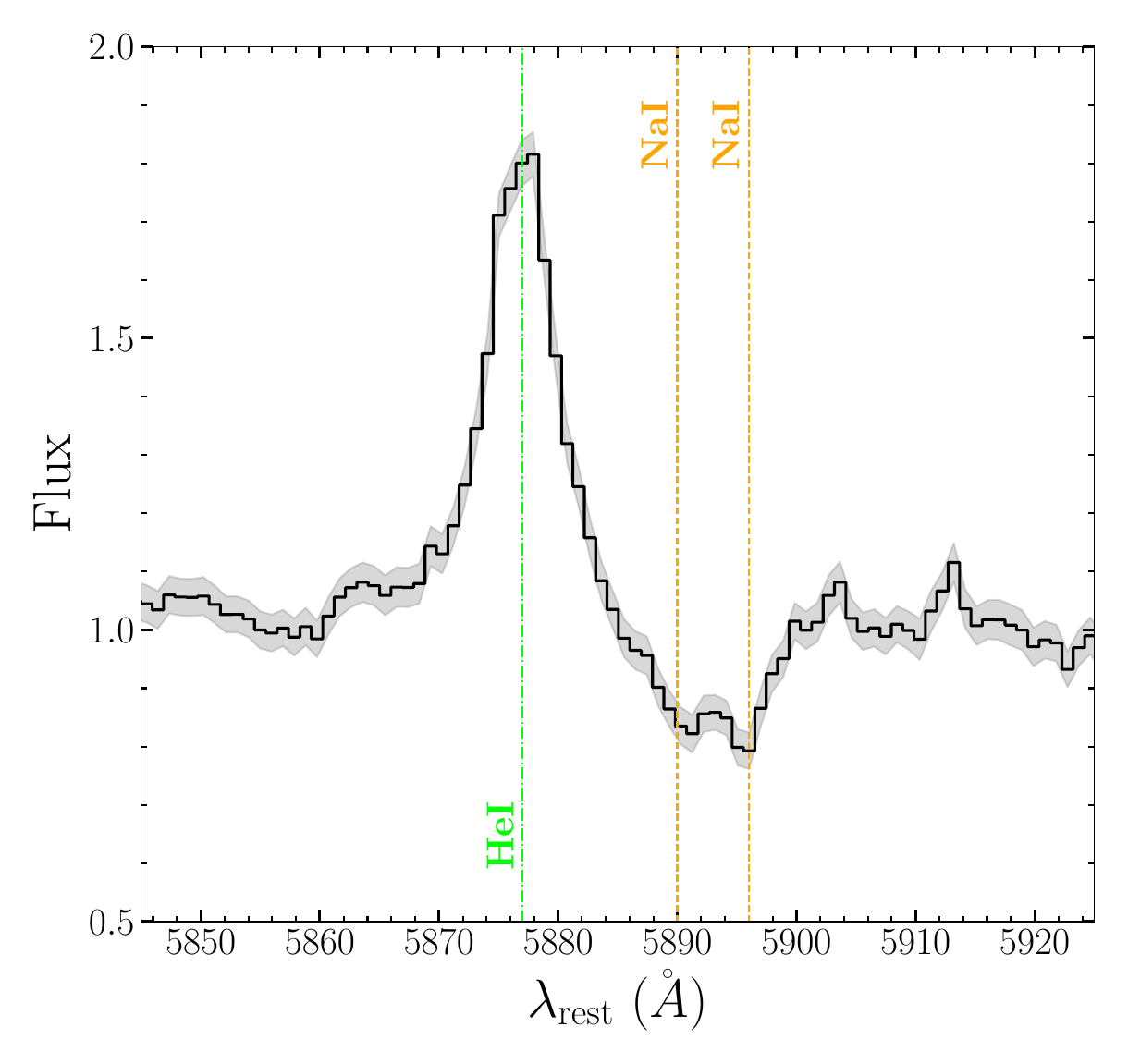}
            \caption{Composite spectrum constructed from the 10 galaxies in our sample that exhibit potential Na D absorption, indicated by the orange dashed lines. The nearby He\,{\sc i} emission line, located to the blue side of Na D, is marked with a lime green dashed line. }
        \label{fig:NaD}
    \end{figure}

    We observe Na\,{\sc i}\,D $\lambda\lambda5890, 5896$ absorption in the G235M grating for 10 galaxies at $1.42\leq z \leq 2.96$. To increase the S/N of the detection, we created a composite spectrum of the 10 galaxies as shown in Figure \ref{fig:NaD}. We compare the average properties of the 10 galaxies with Na\,D absorption to those over the same redshift range with Na\,D coverage but without a significant detection (i.e., 44 galaxies at  $1.38\leq z\leq 2.99$). To assess statistically significant differences, we use a KS test and find that, on average, galaxies with significant Na\,D absorption have higher $M_*$, SFR, and $A_V$ than galaxies without detection of Na\,D absorption. We also measure the outflow velocity of Na\,D using a similar method to the one described in Section \ref{sec:feii_fits}. Due to the blending of the two lines, we fit the two lines simultaneously.
    
   We find that the velocity centroid of the Na\,D absorption profile is consistent with the systemic velocity ($0 \pm 9.7 \text{ km s}^{-1}$), suggesting little to no net outflow in the average population. The lack of a significant blueshift in the composite profile indicates that the outflows may be masked by systemic absorption components. The interpretation of the Na\,D absorption feature is further complicated by substantial stellar contamination since Na\,D absorption is also prominent in the spectra of cool stars. Expanding the sample and increasing the S/N will allow for a more robust exploration of how outflow properties traced by Na\,D vary with galaxy properties at high redshift, enabling direct comparisons with lower-redshift samples.
    
    Previous studies have shown that Na\,D absorption can trace outflows in both local and high-redshift galaxies. In the local universe ($z < 0.3$), blueshifted Na\,D absorption has been reported in infrared-luminous systems, including starburst galaxies \citep{2000Heckman} and both luminous and ultraluminous infrared galaxies \citep{2005Martin, 2009Martin}. \cite{2010Chen} found that the Na\,D absorption profile in local ($z \leq 0.18$) SDSS star-forming galaxies can be split into two components: a quiescent disk-like component that remains near the galaxy's systemic velocity and is stronger in edge on systems, and an outflow component that is most pronounced within $60^\circ$ of the disk's rotation axis (i.e., when galaxies are more face-on). Consistent with our results, they also find Na\,D absorption is stronger in dusty, high-mass, high-SFR galaxies. Similarly, \cite{2022Avery} investigated local, $z\sim0.04$ galaxies, and found that outflows from Na\,D are found in dustier galaxies with elevated star formation. Outside the local universe at $z \sim 2.1$, \cite{2017Casey} found possible Na D absorption in a composite containing 20 galaxies that was blueshifted by $-200 \pm 30 \text{\,km\,s}^{-1}$.

    More recently, there have been detections of Na\,D absorption beyond the local universe  using \textit{JWST}/NIRSpec spectroscopy  \citep{2024Davies,2024Belli,2024DEugenio,2025Sun}. \cite{2024Davies} detected significant blueshifted Na\,D ($\gtrsim100\text{\,km\,s}^{-1}$) in 30 galaxies at $1.7<z<3.5$, predominantly in massive systems ($\log M_*/M_\odot > 10$), but with no strong correlation to dust or SFR, suggesting AGN-driven outflows. Other individual detections at $z \sim 1.3 - 3.1$ also show Na\,D absorption in low-SFR galaxies, where AGN are likely responsible for driving the outflows \citep{2024Belli, 2024DEugenio, 2025Sun}.

    Our sample shows Na\,D absorption in star-forming galaxies with high dust content and elevated SFRs, with host properties that more closely resemble those of local galaxies \citep{2010Chen,2022Avery}. However, the absence of a net blueshift suggests that the outflows may be weaker, more spatially complex, or obscured by significant stellar absorption. These results highlight the need for deeper spectra and larger samples to decompose the Na\,D absorption profile into systemic and outflowing components, and to assess how outflow mechanisms vary across cosmic time and galaxy populations.
 
\section{Conclusions}
Using \textit{JWST}/NIRSpec observations from the AURORA survey, we obtained high-quality near-infrared spectra of star-forming galaxies at $z\gtrsim 2$, enabling the use of rest-frame NUV absorption line diagnostics at these high redshifts. We measured the outflow velocities from Fe\,{\sc ii} and Mg\,{\sc ii} absorption features in 41 and 43 galaxies, respectively, and investigated how these velocities and Mg\,{\sc ii} emission correlate with stellar and dust properties, including $M_*$, SFR, sSFR, $\Sigma_{\text{SFR}}$, $\beta$, $E(B-V)$, and $A_V$, using both individual and composite spectra. Our main results are summarized as follows:

    \begin{enumerate}
        \item Outflow velocities from individual galaxies show no significant correlation with galaxy properties. However, sample distributions demonstrate that galaxies with detected outflows have higher $M_*$ than galaxies without detections, suggesting underlying trends may be masked by sample limitations such as a narrow dynamic range in galaxy properties. Additionally, composite spectra reveal that $v_{\text{max}}$ increases with $M_*$, SFR, and dust properties.
        \item Galaxies exhibiting Mg\,{\sc ii} emission show lower $M_*$, higher sSFR, more negative $\beta$, and lower $A_V$ compared to non-emitters, consistent with trends observed at lower redshift. The composite spectra also show similar trends where galaxies with higher $M_*$, $\beta$, and $A_V$ have stronger Mg\,{\sc ii} absorption. These findings suggest that Mg\,{\sc ii} emission is more prevalent in low-mass, actively star-forming galaxies with less obscured stellar populations.
        \item \textit{JWST} facilitates the direct, simultaneous analysis of NUV and FUV absorption features within individual galaxy spectra. In our sample, one galaxy exhibits both NUV and FUV transitions with measured outflow velocities that are consistent within $2\sigma$. Expanding this subset in future studies will enable a more comprehensive comparison of the kinematics traced by NUV and FUV lines.
        \item We detect Na D absorption in 10 galaxies at $1.42 \leq z \leq 2.96$ in our sample. Comparing this sample to 44 galaxies at $1.38 \leq z \leq 2.99$ without Na D absorption, we find that those with Na D absorption tend to have higher $M_*$, SFR, and $A_V$. This suggests that Na D absorption is more prevalent in more massive, actively star-forming, and dustier galaxies.
    \end{enumerate}

\textit{JWST} has enabled the first use of NUV absorption-line diagnostics to study galactic outflows at $z\gtrsim2.5$, filling a crucial gap in our understanding of how outflows regulate galaxy evolution across cosmic time. While individual outflow velocities do not show correlations with galaxy properties, composite spectra reveal trends linking outflows to stellar mass, star formation, and dust content. The presence of Mg\,{\sc ii} emission and Na D absorption in our sample highlights the diverse characteristics of outflows associated with different galaxy properties. Importantly, the ability to use both NUV and FUV tracers within the same spectrum allows for direct comparisons of outflow kinematics using multiple diagnostics. These findings highlight the importance of using consistent tracers across redshifts to understand how outflows influence star formation and chemical enrichment. Future \textit{JWST} surveys will be key to deepening this understanding.

\begin{acknowledgments}
This work is based on observations made with the NASA/ESA/CSA James Webb Space Telescope. The data were
obtained from the Mikulski Archive for Space Telescopes at
the Space Telescope Science Institute, which is operated by the
Association of Universities for Research in Astronomy, Inc.,
under NASA contract NAS5-03127 for JWST.  The specific observations analyzed can be accessed via \dataset[DOI: 10.17909/6mza-5q55]{https://archive.stsci.edu/doi/resolve/resolve.html?doi=10.17909/6mza-5q55}.
We also acknowledge support from NASA grant JWST-GO-01914. FC acknowledges support from a UKRI Frontier Research Guarantee Grant (PI Cullen; grant reference: EP/X021025/1).
DJM acknowledges the support of the Science and Technology Facilities Council. PO acknowledges the Swiss State Secretariat for Education, Research and Innovation (SERI) under contract number MB22.00072, as well as from the Swiss National Science Foundation (SNSF) through project grant 200020$\_$207349. AJP was generously supported by a Carnegie Fellowship through the Carnegie Observatories. DN was funded by JWST-AR-01883.001.
\end{acknowledgments}

%


\bibliography{ms}{}
\bibliographystyle{aasjournal}



\end{document}